\documentstyle[preprint,aps,psfig]{revtex}
\tightenlines
\newtheorem{theorem}{Theorem}[section]
\newtheorem{lemma}{Lemma}[section]

\begin{document}
\title{Amplitude Equations for Electrostatic Waves: multiple species}
\author{John David Crawford and Anandhan Jayaraman}
\address{Department of Physics and Astronomy\\
University of Pittsburgh\\
Pittsburgh, Pennsylvania  15260}
\date{\today}
\maketitle

\begin{abstract}
The amplitude equation for an unstable electrostatic wave is analyzed using
an expansion in
the mode amplitude $A(t)$. In the limit of weak instability, i.e.
$\gamma\rightarrow0^+$ where
$\gamma$
is the linear growth rate, the nonlinear coefficients are singular and their
singularities predict the
dependence of $A(t)$ on $\gamma$. Generically the scaling
$|A(t)|=\gamma^{5/2}r(\gamma t)$ as $\gamma\rightarrow0^+$ is required to
cancel the coefficient  singularities to all orders.
This result predicts
the electric field scaling $|E_k|\sim\gamma^{5/2}$ will
hold universally for these instabilities (including beam-plasma and two-stream
configurations) throughout the dynamical evolution and in the
time-asymptotic state. In exceptional cases, such as infinitely massive ions,
the coefficients are less singular and the more familiar trapping scaling
$|E_k|\sim\gamma^2$ is recovered.

\end{abstract}

%-------------------------------------------------------
\section{Introduction}
%------------------------------------------------------
\subsection{Overview and background}
We recently presented a detailed analysis of the amplitude equation for
an unstable electrostatic mode in an unmagnetized Vlasov plasma
(henceforth (I)).\cite{jdc95}
In that work,  the amplitude equation was
studied using an expansion in the wave amplitude, and only the electron
motion was considered with
the ions treated as a fixed neutralizing background.
 In this paper we study the effect of including mobile ions and investigate the
mode amplitude equation for an
electrostatic wave in a multi-species unmagnetized plasma. As in (I), we
focus on the regime of weak instability and carefully analyze the structure
of the amplitude equation in the the limit $\gamma\rightarrow0^+$
where $\gamma$ is the linear growth rate for the unstable wave.

Incorporating the effects of finite ion inertia is a significant step for
several reasons. With finite mass, the ion motion occurs on a finite
timescale $T_i<\infty$ and in the limit of very weak growth rates
this will be a fast timescale, i.e. $T_i<<\gamma^{-1}$ as
$\gamma\rightarrow0^+$. Over the duration of the instability, the ions
can readily respond to the electric field of the wave especially resonant
ions near the phase velocity. This ion response is always present even for high
frequency modes such as plasma waves since at the phase velocity the wave
frequency is Doppler-shifted to zero. In addition, there are modes such
as ion-acoustic waves that depend on ion density oscillations and are
suppressed when the ions are treated as fixed. Allowing for
mobile ions is a prerequisite, if we seek the amplitude equation
for an unstable ion-acoustic mode.

This
paper can be read independently, but the reader is referred  to (I) for a
discussion of the history of the problem and additional
introductory material.
An essential difference between our approach
and previous work lies in the choice of unperturbed
state. Earlier theories assumed an equilibrium with a neutrally
stable mode and obtained ill-defined expansion coefficients.
\cite{frieman}-\cite{burnap}
This can be
avoided by
taking the weakly unstable equilibrium as the unperturbed
state; a choice that
naturally leads one to work with the unstable manifold.
A brief summary of our results on the multi-species case has appeared
elsewhere.\cite{jdcaj}

In the limit $\gamma\rightarrow0^+$, the amplitude equation defines a kind
of singular perturbation problem whose detailed features reveal asymptotic
scaling behavior of the nonlinear wave. This is a key idea behind our approach
and it can be formulated more precisely as follows.  The mode eigenvalue
$\lambda=\gamma-i\omega$ can be complex (beam-plasma) or real (two-stream), but
in either case the equations for the amplitude $A(t)=\rho(t)\,e^{-i\theta(t)}$
have the form
\begin{eqnarray}
\dot{\rho}&=&\gamma\rho+a_1\rho^3+a_2\rho^5+{\cal
O}({\rho^7})\label{eq:hopf1}\\
\dot{\theta}&=&\omega+{a'_1}\rho^2+{a'_2}\rho^4+{\cal
O}({\rho^6}).\label{eq:hopf2}
\end{eqnarray}
Since the Vlasov equilibrium is assumed to be spatially homogeneous,  the
evolution of $\rho(t)$ decouples from the phase $\theta(t)$ . Suppose for
simplicity that the coefficients $a_j$  are constant (ignoring
their dependence on $\gamma$), then we may view (\ref{eq:hopf1})
as a singular perturbation problem with the linear term
$\gamma\,\rho$ representing the perturbation.
If $\gamma>0$, then there is always
a neighborhood of the equilibrium $\rho=0$ where the perturbation dominates
the unperturbed system and can completely change the dynamics.

As in other such singular problems, a possible strategy is to seek a (singular)
change of variables which transforms (\ref{eq:hopf1}) into a regular
perturbation problem. Thus it is natural to define a new amplitude $r(\tau)$
by
\begin{equation}
{\rho}(t)=\gamma^\beta r(\gamma^\delta t)\label{eq:newvar}
\end{equation}
and rewrite (\ref{eq:hopf1}) in terms of $r(\tau)$
\begin{equation}
\frac{d r}{d\tau} =\gamma^{1-\delta}\,r(\tau) + \gamma^{2\beta-\delta}
a_1(\gamma)r^3+\gamma^{4\beta-\delta}a_2(\gamma)r^5+\cdots.\label{eq:aeqnb}
\end{equation}
If possible, the choice of $\beta$ and $\delta$ should be made so that
({\em i}) each term in (\ref{eq:aeqnb}) is well behaved as
$\gamma\rightarrow0^+$
and ({\em ii}) the effect of $\gamma>0$ is a regular perturbation
of the system at $\gamma=0$. The latter condition requires $\delta=1$;
otherwise the linear term will continue to be
a singular perturbation. The exponent $\beta$ then needs to be chosen (if
possible)
to achieve a balance between the nonlinear terms and the linear term.
In
the simplest situation, when the coefficients
$a_1, a_2,\ldots$ have
well-defined {\em finite} limits  as $\gamma\rightarrow0$, then one has
the exponent $\beta=1/2$ and this yields
\begin{equation}
\frac{d r}{d\tau} =r(\tau) + a_1(\gamma)r^3 +\gamma\,a_2(\gamma)r^5+\cdots.
\label{aeqnc}
\end{equation}
The unperturbed system $\dot{r}=r+a_1(0)r^3$ now includes the linear term
and near $r=0$ the effect of small $\gamma$ is unimportant. The change of
variables (\ref{eq:newvar}) is singular at $\gamma=0$ as expected, but
nevertheless we have determined the overall scaling $\beta=1/2$ for nonlinear
solutions
near the equilibrium in the regime of weak instability. In addition, since
terms in (\ref{eq:aeqnb}) at fifth order or higher are at least of order
${\cal O}(\gamma)$, the unperturbed problem truncates to a simple balance
between
the linear instability and the dominant nonlinear term.

For an unstable electrostatic wave, the situation is quite different because
the nonlinear coefficients $a_j$ blow up as $\gamma\rightarrow0$ and
the exponent $\beta=1/2$ does not yield a well-behaved equation for $r(\tau)$.
We attack the problem of finding a satisfactory value for $\beta$
in two steps. First, we determine what choice for $\beta$ will control
the singularity of the cubic coefficient $a_1$ in (\ref{eq:hopf1}), and
then we investigate whether this
choice will also  remove the singularities of the higher
order nonlinear terms. Typically, the cubic coefficient diverges
asymptotically like $\gamma^{-4}$ in which case we must choose
$\beta\geq5/2$ in order to obtain a finite cubic
term in the rescaled amplitude equation (\ref{eq:aeqnb}).

Our analysis  shows that
setting $\beta=5/2$ yields a theory that is finite to all orders as
$\gamma\rightarrow0$ at $\gamma=0$ retains (rescaled) nonlinear terms that
balance the linear instability.
Choosing a larger exponent $\beta=5/2+\epsilon$ would also  yield a finite
theory to all orders,
but now the cubic term is of order ${\cal O}(\gamma^{2\epsilon})$ and all the
higher order terms ($j>1$) are of order ${\cal O}(\gamma^{2\epsilon j})$;
there
is no balance between the linear and nonlinear terms at $\gamma=0$.
For $\epsilon>0$, the larger exponent suppresses nonlinear effects too
strongly.
In this sense, the scaling
\begin{equation}
{\rho}(t)=\gamma^{5/2} r(\gamma t)\label{genscale}
\end{equation}
is uniquely determined. Of the two steps, the second is by far the most
difficult; it
is relatively straightforward to compute $a_1$ and determine its divergence
(Section \ref{sec:pinch}). Verifying that
a given choice of $\beta$ works to all orders requires a detailed study of
the recursion relations of the theory and the final theorem is proved by
induction
(Section \ref{sec:all orders}).

In special cases, the cubic coefficient is less
singular, diverging like $\gamma^{-3}$ instead of $\gamma^{-4}$. The limit of
infinitely massive ions
is the best known example of this situation, and in (I) we  found that
$\beta=2$ was the scaling needed to make the nonlinear amplitude equation
finite to all orders. In this paper, we identify two additional
examples where the cubic coefficient has a $\gamma^{-3}$ singularity: if the
effects of the resonant ions are negligible (in a sense we make precise in
Section \ref{subsec:special}) or if the electron and ion masses are equal, i.e.
for an electron-positron plasma. In the former
case we prove that the exponent $\beta=2$ is correct, but the precise value
of $\beta$ for the electron-positron system is not unambiguously determined due
special complications in the singularity structure of the
amplitude expansion.

\subsection{Notation}

Our model is a one-dimensional, multi-species Vlasov plasma defined
by
\begin{equation}
\frac{\partial F^{(s)}}{\partial t}+v\frac{\partial F^{(s)}}{\partial x}+
\kappa^{(s)}\;E\;\frac{\partial F^{(s)}}{\partial v}=0\label{eq:vlasov}
\end{equation}
\begin{equation}
\frac{\partial E}{\partial
x}=\sum_s\int^\infty_{-\infty}\,dv\,F^{(s)}(x,v,t).\label{eq:poisson}
\end{equation}
Here $x$, $t$ and $v$ are measured in units of $u/\omega_e$, $\omega_e^{-1}$
and $u$, respectively, where $u$ is a chosen velocity scale and
$\omega_e^2=4\pi e^2n_e/m_e$.
The plasma length is $L$ with periodic boundary conditions and we adopt the
normalization
\begin{equation}
\int^{L/2}_{-L/2}\,dx\,\int^\infty_{-\infty}\,dv\,F^{(s)}(x,v,t)=
\left(\frac{z_s\;n_s}{n_e}\right)L\label{eq:Fnorm}
\end{equation}
where $q_s=e\,z_s$ is the charge of species $s$ and $\kappa^{(s)}\equiv
{q_sm_e/em_s}$. Note that $\kappa^{(e)}=-1$ for electrons and
that the normalization (\ref{eq:Fnorm}) for negative species
makes the distribution function negative.

Let $F_0(v)$ and $f(x,v,t)$ denote the multi-component fields for the
equilibrium and
perturbation respectively and $\kappa$ the matrix of mass ratios,
\begin{equation}
 f\equiv\left(\begin{array}{c}
f^{(s_1)}\\f^{(s_2)}\\ \vdots\end{array}\right)\;\;\;\;
F_0\equiv\left(\begin{array}{c}F_0^{(s_1)}\\F_0^{(s_2)}\\
\vdots\end{array}\right)\;\;\;\;
\kappa\equiv\left(\begin{array}{cccc}\kappa^{(s_1)} & 0 & 0 & \cdots\\
 0&\kappa^{(s_2)}&0&\cdots\\ \vdots&\vdots&\vdots\end{array}\right),
\end{equation}
then the system (\ref{eq:vlasov}) - (\ref{eq:poisson}) can be concisely
expressed as
\begin{equation}
\frac{\partial f}{\partial t}={\cal{L}}\,f+{\cal{N}}(f)\label{eq:dynsys}
\end{equation}
where the linear operator is defined by
\begin{eqnarray}
{\cal{L}}\,f&=&\sum^{\infty}_{l=-\infty}\,e^{ilx}\,(L_l f_l)(v)
\label{eq:fexp}\\
(L_l f_l)(v)&=&\left\{\begin{array}{cc}0&l=0\\
-il\left[vf_l(v)+\kappa\cdot\eta_l(v)\sum_{s'}
\int^\infty_{-\infty}\,dv'\,f^{(s')}_l(v')
\right]&l\neq0,\end{array}\right.\label{eq:linop}
\end{eqnarray}
with $\eta_l(v)\equiv-\partial_vF_0/l^2$,
and the nonlinear operator ${\cal{N}}$ is
\begin{equation}
{\cal{N}}(f)=\sum^{\infty}_{m=-\infty}\,
e^{imx}\,{\sum^{\infty}_{l=-\infty}}'\, \frac{i}{l}\left(\kappa\cdot
\frac{\partial
f_{m-l}}{\partial v}\right)
\sum_{s'}\int^\infty_{-\infty}\,dv'\,f^{(s')}_l(v').
\label{eq:nop}
\end{equation}
In the spatial Fourier expansion (\ref{eq:fexp}), $l$ denotes an integer
multiple of the basic wavenumber $2\pi/L$,  and a primed summation as in
(\ref{eq:nop}) omits the $l=0$ term. The notation $\kappa\cdot\eta_l(v)$ or
$\kappa\cdot\partial_v f_{m-l}$
denotes matrix multiplication.

An equilibrium $F_0(v)$ depends on parameters such as the densities or
temperatures of the various species, but this dependence will be suppressed
unless it is explicitly needed.

Symmetries of the model (\ref{eq:dynsys}) and the
equilibrium $F_0(v)$ are important qualitative features of the problem,
and play a significant role in Section \ref{sec:expand} when we formulate
the amplitude expansions.
Spatial translation,
${\cal T}_a:(x,v)\rightarrow (x+a,v),$
and reflection,
${{\cal R}:(x,v)\rightarrow (-x,-v)},$
act as operators on $f(x,v,t)$ in the usual way: if $\alpha$
denotes an arbitrary transformation then $(\alpha\;f)(x,v)\equiv
f(\alpha^{-1}\,(x,v))$ where $(\alpha\;f)(x,v)$
denotes the transformed distribution
function. The operators ${\cal{L}}$ and ${\cal{N}}$ commute
with ${\cal T}_a$ due to the spatial homogeneity of $F_0$,
and if $F_0(v)=F_0(-v)$,
then ${\cal{L}}$ and ${\cal{N}}$ also commute with the reflection operator
${\cal R}$.
With periodic boundary conditions, $x$ is a periodic coordinate so
${\cal T}_a$ and ${\cal R}$ generate ${\rm O}(2)$, the symmetry group of the
circle. If only the translation symmetry is present, then the group is ${\rm
SO}(2)$.

An inner product is needed in Section \ref{sec:expand} to derive the amplitude
equation. For two multi-component fields of $(x,v)$, e.g.
$B=(B^{(s_1)},B^{(s_2)},B^{(s_3)},..)$ and
$D=(D^{(s_1)},D^{(s_2)},D^{(s_3)},..)$, we define their inner product by
\begin{eqnarray}
(B,D)&\equiv&\sum_s\int^{L/2}_{-L/2}\,dx\int_{-\infty}^{\infty}\,dv\,
B^{(s)}(x,v)^\ast D^{(s)}(x,v)
=\int^{L/2}_{-L/2}\,dx\,<B,D>
\end{eqnarray}
where
\begin{eqnarray}
<B,D>&\equiv&\sum_s \int_{-\infty}^{\infty}\,dv\,B^{(s)}(x,v)^\ast
D^{(s)}(x,v).
\end{eqnarray}

\subsection{Summary of linear theory}

The spectral theory for ${\cal{L}}$ is well established, and the needed results
are simply recalled to establish our notation.\cite{vkamp,case,cra1} The
eigenvalues $\lambda=-il z$ of ${\cal{L}}$ are determined by the roots
$\Lambda_{l}(z)=0$ of the ``spectral function'',
\begin{equation}
\Lambda_{l}(z)\equiv 1+
\int^\infty_{-\infty}\,dv\,\frac{\sum_s\kappa^{(s)}\eta_l^{(s)}(v)}{v-z}.
\label{eq:specfcn}
\end{equation}
 If the
contour in (\ref{eq:specfcn}) is replaced by the Landau contour for ${\rm
Im}(z)<0$ then
we have the linear dielectric $\epsilon_{{l}}(z)$; for ${\rm Im}(z)>0$,
$\Lambda_{l}(z)$ and $\epsilon_{{l}}(z)$ are the same function.
 For ${\rm SO}(2)$
symmetry, the eigenvalues are generically complex; for ${\rm O}(2)$
symmetry,
when the equilibrium is also reflection-symmetric, then real eigenvalues can
occur as in a two-stream instability; see for example \cite{cowley}.

Associated with an eigenvalue $\lambda=-il z$ is the multi-component
eigenfunction $\Psi(x,v)=e^{ilx}\,\psi(v)$ where
\begin{equation}
\psi(v)=-\frac{\kappa\cdot \eta_l}{v-z}.%\label{eq:efcn2}
\end{equation}
There is also an associated adjoint eigenfunction
$\tilde{\Psi}(x,v)=e^{ilx}\tilde{\psi}(v)/L$
satisfying $(\tilde{\Psi},\Psi)=1$ with
\begin{equation}
\tilde{\psi}(v)=- \frac{1}{\Lambda'_{l}(z)^\ast(v-z^\ast)}.
\label{eq:aefcn2}
\end{equation}
Note that all components of $\tilde{\psi}(v)$ are the same.
The normalization in (\ref{eq:aefcn2}) assumes that the root of
$\Lambda_{l}(z)$ is simple and is chosen so that $<\tilde{\psi},\psi>=1$.
The adjoint determines the projection of $f(x,v,t)$ onto the eigenvector, and
this projection defines the time-dependent amplitude of $\Psi$, i.e.
$A(t)\equiv(\tilde{\Psi},f)$.

%_________________________________________________________________

\section{Amplitude equation on the unstable manifold}\label{sec:expand}
%___________________________________________________________________

%----------------------------------------------------------
\subsection{Unstable linear modes}
%----------------------------------------------------------

The equilibrium $F_0(v)$ is assumed to support a ``single'' unstable mode in
the following sense. We shall assume that $E^u$, the unstable subspace
for ${\cal L}$, is two-dimensional. With translation symmetry and periodic
boundary conditions, this is the
simplest instability problem that can be posed.

Henceforth, let $k$ denote the wavenumber of this unstable mode that is
associated with the root $\Lambda_{k}(z_0)=0$ which we assume to be simple,
i.e. $\Lambda'_{k}(z_0)\neq0.$
The corresponding
eigenvector is
\begin{equation}
\Psi_1(x,v)=e^{ikx}\,\psi(v)=
e^{ikx}\left(-\frac{\kappa\cdot \eta_k}{v-z_0}\right).
\label{eq:lefcn}
\end{equation}
The root  $z_0=v_p+{i\gamma}/{k}$ determines the phase velocity $v_p=\omega/k$
and the growth rate $\gamma$  of the linear mode as the real and imaginary
parts of the eigenvalue $\lambda=-ikz_0=\gamma-i\omega$.
{}From (\ref{eq:specfcn}) and $\Lambda_{k}(z_0)=0$, we obtain
\begin{equation}
\Lambda_{l}(z_0)=(l^2-k^2)/{l^2};\label{eq:specfcnid}
\end{equation}
this identity will be useful shortly.

When $F_0(v)$ lacks reflection symmetry,
then $\Psi_1$ typically has a
non-zero phase velocity and $\lambda$ is complex. In this case, the identities
$\Lambda_{k}(z)=\Lambda_{-k}(z)$ and $\Lambda_{k}(z)^\ast=\Lambda_{k}(z^\ast)$
imply three additional modes: $\Psi_1^\ast$, $\Psi_2$, and $\Psi_2^\ast$ where
\begin{equation}
\Psi_2(x,v)=e^{ikx}\,\psi(v)^\ast.
\end{equation}
These eigenfunctions correspond to eigenvalues $\lambda^\ast$, $-\lambda^\ast$,
and $-\lambda$, respectively, and fill out the eigenvalue quartet
characteristic of Hamiltonian systems. In the absence of reflection symmetry,
the eigenvalues are typically simple and the unstable subspace is
two-dimensional.

When $F_0(v)$ is reflection-symmetric, the
eigenvalues may be either real or complex. In either case $\Psi_j$
and ${\cal{R}}\cdot\Psi_j$ are linearly independent so the eigenvalues have
multiplicity two. Now $E^u$ is two-dimensional only in the case of a real
eigenvalue since a multiplicity-two complex conjugate pair implies a
four-dimensional
unstable subspace; this latter possibility is not considered further.

The components of the distribution function
along the unstable eigenvectors $\Psi_1$ and $\Psi_1^\ast$ are identified by
writing
\begin{equation}
f(x,v,t)=\left[A(t)\Psi(x,v) + cc\right] + S(x,v,t)\label{eq:linmodes}
\end{equation}
where $A(t)=(\tilde{\Psi},f)$ is the mode amplitude for $\Psi$ and
$(\tilde{\Psi},S)=0$.
In (\ref{eq:linmodes}), the subscript
on $\Psi_1$ has been dropped, and $\tilde{\Psi}=\exp(ikx)\,\tilde{\psi}/L$
is the adjoint function for $z_0$ from (\ref{eq:aefcn2}).

The action of
translation ${\cal T}_a$ and reflection ${\cal{R}}$ on $f(x,v,t)$
implies an action by these operators on the variables $(A,S)$.  From
(\ref{eq:linmodes}) we note that ${\cal T}_af(x,v,t)=f(x-a,v,t)$
is equivalent to
\begin{equation}
{\cal T}_a \,(A,S(x,v))=(e^{-ik a}A, S(x-a,v)),\label{eq:Atrans}
\end{equation}
and reflection ${\cal{R}}f(x,v,t)=f(-x,-v,t)$ yields
\begin{equation}
{\cal{R}}\, (A,S(x,v))=(A^\ast,S(-x,-v)).\label{eq:Aref}
\end{equation}
When these transformations are symmetries, these relations are important for
organizing the amplitude expansions introduced in Section \ref{sec:pinch}.

%--------------------------------------------------------------
\subsection{Derivation of the amplitude equation}
%-------------------------------------------------------------

In the  $(A,S)$ variables, the Vlasov equation (\ref{eq:dynsys}) becomes:
\begin{eqnarray}
\dot{A}&=&\lambda\,A+(\tilde{\Psi},{\cal{N}}(f))\label{eq:adot}\\
\frac{\partial S}{\partial t}&=&{\cal{L}}
S+{\cal{N}}(f)-\left[(\tilde{\Psi},{\cal{N}}(f))\,\Psi +
cc\right]\label{eq:Sdot}
\end{eqnarray}
where
\begin{equation}
(\tilde{\Psi},{\cal{N}}(f))=-{i}\,{\sum^{\infty}_{l=-\infty}}'
\frac{1}{l}\;
<{\partial_v}\,\tilde{\psi},  \kappa\cdot  f_{k-l}>
\sum_{s'}\;\int^\infty_{-\infty}\,dv'\,f^{(s')}_l(v').\label{eq:projnl}
\end{equation}
In writing (\ref{eq:adot}) we have used the adjoint relationship
$(\tilde{\Psi},{\cal{L}} S)=
({\cal{L}}^\dagger\tilde{\Psi},S)=\lambda(\tilde{\Psi},S)=0$ and in
(\ref{eq:projnl}) an integration by parts shifts the velocity
derivative onto $\tilde{\psi}$. These coupled equations are equivalent to
(\ref{eq:dynsys}); however by restricting them to the unstable manifold we
obtain an autonomous equation for $A(t)$. This procedure is briefly summarized;
a more detailed description is provided in (I).

The unstable subspace $E^u$ is invariant under the linear dynamics $\partial_t
f={\cal{L}} f$, but this no longer holds once the nonlinear terms ${\cal{N}}
(f)$ are included. Instead we assume there is a two-dimensional invariant
surface, the unstable manifold $W^u$, that represents
the nonlinear deformation
of $E^u$. The unstable manifold is tangent to $E^u$ at the equilibrium, and
near $F_0$ it can be described by a function $H(x,v,A,A^\ast)$ which locates
the
manifold ``above'' the point $[A\Psi +A^\ast\Psi^\ast]$ in $E^u$; the
geometry is illustrated in Fig. 1. \\

\centerline{\psfig{file=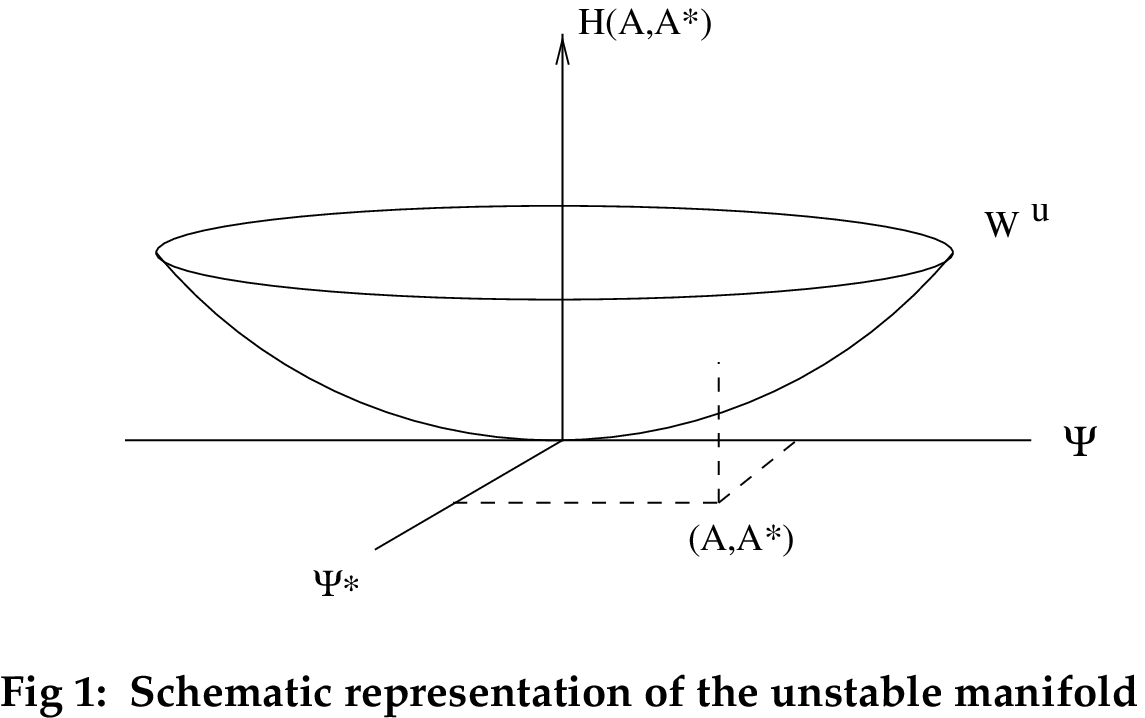,width=4.5in}}

Thus
\begin{equation}
f^u(x,v,t)=\left[A(t)\Psi(x,v) + cc\right] +
H(x,v,A(t),A^\ast(t))\label{eq:fu}
\end{equation}
represents a distribution function on $W^u$.
In this expression, the evolution of $S$ be determined from $H$, i.e.
\begin{equation}
S(x,v,t)=H(x,v,A(t),A^\ast(t))=\left(
\begin{array}{c}H^{(s_1)}(x,v,A(t),A^\ast(t))\\
H^{(s_2)}(x,v,A(t),A^\ast(t))\\
\vdots
 \end{array}\right).\label{eq:Su}
\end{equation}
When this representation is
substituted into (\ref{eq:adot}) - (\ref{eq:Sdot}) we obtain
\begin{eqnarray}
\dot{A}&=&\lambda\,A+(\tilde{\Psi},{\cal{N}}(f^u))\label{eq:aeqn}\\
\left.\frac{\partial S}{\partial t}\right|_{f^u} &=&{\cal{L}}
H+{\cal{N}}(f^u)-\left[(\tilde{\Psi},{\cal{N}}(f^u))\,\Psi +
cc\right].\label{eq:Sdotwu}
\end{eqnarray}
Note that (\ref{eq:aeqn}) defines an {\em autonomous} two-dimensional flow
describing
the self-consistent nonlinear evolution of the unstable mode; this is the
amplitude equation we shall study.

Certain general features of the amplitude equation follow from the
transformations (\ref{eq:Atrans}) - (\ref{eq:Aref}) when these are symmetries
of the Vlasov equation (\ref{eq:dynsys}). For
our problem, the amplitude equation always has translation symmetry
(\ref{eq:Atrans}) and we can apply standard results on the form of such
symmetric equations. In particular, a two-dimensional vector field
$\dot{A}=V(A,A^\ast)$ that is symmetric with respect to $A\rightarrow
A\,e^{-ika}$ can be written as $V(A,A^\ast)=A\,p(\sigma)$ where $\sigma=|A|^2$
and $p(\sigma)$ is a smooth function determined from $V$.~\cite{gss}
Hence we know the right hand side of (\ref{eq:aeqn}) takes the general form,
\begin{equation}
A\,p(\sigma)=\lambda\,A+(\tilde{\Psi},{\cal{N}}(f^u)),
\label{eqn:pdef}
\end{equation}
where the function $p(\sigma)$ must be determined from the Vlasov
equation. Typically $p(\sigma)$ is
complex-valued, however when
$F_0$ is reflection-symmetric then $p(\sigma)$ is forced to be real.

%---------------------------------------------------
\subsection{Analysis of $H(x,v,A,A^\ast)$}
%---------------------------------------------------

An equation for $H$ follows by requiring consistency between
(\ref{eq:Su}) and (\ref{eq:Sdotwu}). Equating the time derivative of
(\ref{eq:Su}) with the right hand side of (\ref{eq:Sdotwu}) gives
\begin{equation}
\frac{\partial H}{\partial A}\,\dot{A}+\frac{\partial H}{\partial
A^\ast}\,\dot{A}^\ast = {\cal{L}}\,H+
{\cal{N}}(f^u)-\left[(\tilde{\Psi},{\cal{N}}(f^u))\,\Psi + cc\right]
\label{eq:Heqn}
\end{equation}
which is to be solved for $H$ subject to (\ref{eq:tang}).
Since the manifold is tangent to $E^u$ at the equilibrium, $H$ must
satisfy
\begin{equation}
0=H(x,v,0,0)=\frac{\partial H}{\partial A}(x,v,0,0)=
\frac{\partial H}{\partial A^\ast}(x,v,0,0).\label{eq:tang}
\end{equation}

The symmetries of the problem impose general restrictions on $H$ because
the unstable manifold is mapped onto itself by a symmetry transformation.
For a translation this invariance means that  $f^u(x-a,v,t)$  corresponds to a
point on the manifold and thus can be written as,
\begin{eqnarray}
\left({\cal T}_a f^u \right)(x,v)&=&
[A\Psi(x-a,v) +cc] + H(x-a,v,A,A^\ast)\\
&=&[A'\Psi(x,v) +cc] + H(x,v,A',A'^\ast).
\end{eqnarray}
Consistency between these two expressions requires that the transformed
amplitude $A'$ is related to the initial
amplitude $A$ by $A'=e^{-ika} A$ (cf. (\ref{eq:Atrans})), and this implies
that $H$ satisfies
\begin{equation}
H(x-a,v,A,A^\ast)=H(x,v,e^{ika}A,e^{-ika}A^\ast).\label{eq:Htrans}
\end{equation}
If, in addition, the system  (\ref{eq:adot}) - (\ref{eq:Sdot}) has reflection
symmetry, then  similar reasoning yields
\begin{equation}
H(-x,-v,A,A^\ast)= H(x,v,A^\ast,A).
\end{equation}

These general symmetry properties of $H$ usefully
constrain the form of the Fourier expansion
\begin{equation}
H(x,v,A,A^\ast)=\sum^{\infty}_{l=-\infty}\,e^{ilx}\,H_l(v,A,A^\ast)
=\sum^{\infty}_{l=-\infty}\,e^{ilx}\,\left(
\begin{array}{c}H^{(s_1)}_l(v,A,A^\ast)\\
H^{(s_2)}_l(v,A,A^\ast)\\
\vdots
 \end{array}\right).
\label{eq:Hexpand}
\end{equation}
Applying (\ref{eq:Htrans}) shows that the components $H_l$ must vanish unless
$l$ is an integer multiple of $k$ and that the non-zero components have the
general form
\begin{eqnarray}
H_0(v,A,A^\ast)&=&\sigma\,h_0(v,\sigma)\nonumber\\
H_k(v,A,A^\ast)&=&A\sigma\,h_1(v,\sigma)\label{eq:hdef}\\
H_{mk}(v,A,A^\ast)&=&A^m\,h_m(v,\sigma)\;\;\;\;{\mbox{for}}\;\;m\geq2\nonumber
\end{eqnarray}
where $H_{-l}={H_l}^\ast$. The functions $h_m$ are not determined by symmetry;
however, if reflection symmetry also holds, then
\begin{equation}
h_m(-v,\sigma)=h_m(v,\sigma)^\ast.
\end{equation}

The reasoning leading to these symmetry results can be briefly outlined.
The translation symmetry (\ref{eq:Htrans}) requires
\begin{equation}
e^{-ila} H_l(v,A,A^\ast) = H_l(v,A e^{-ika}, A^\ast e^{ika}).
\label{eq:Hftrans}
\end{equation}
By choosing $a=2\pi/k$, this implies we find $H_l=0$ unless $\exp(2\pi i
(l/k))=1$, hence the non-zero components satisfy $l=mk$ for integer $m$.
Now recall that a function $G(v,A,A^\ast)$ which is invariant under
$A\rightarrow Ae^{-ika}$ has the form
$G(v,A,A^\ast)=g(v,\sigma)$ where $\sigma=|A|^2$ and $g(v,\sigma)$ is a smooth
function
determined by $G$. The relation in (\ref{eq:Hftrans}) implies that
$G(v,A,A^\ast)= (A^\ast)^m H_{mk}(v,A,A^\ast)$ is such a invariant function.
The factor $(A^\ast)^m$ in $G$ requires that the corresponding $g(v,\sigma)$
contain an overall factor of $\sigma^m$; hence we set $g(v,\sigma)=\sigma^m
h(v,\sigma)$. The resulting expression, $(A^\ast)^m H_{mk} = \sigma^m
h(v,\sigma)$, then implies
$H_{mk} = A^m h(v,\sigma)$. This general form, together with the requirement
(\ref{eq:tang}), leads to general expressions for the Fourier components of $H$
in (\ref{eq:hdef}).

%-------------------------------------------------------------
\subsection{Equations for $p(\sigma)$ and $h_m(v,\sigma)$}
%-------------------------------------------------------------

The calculation of the amplitude equation (\ref{eq:aeqn}) reduces to a
determination of the function $p(\sigma)$ in (\ref{eqn:pdef}); this requires
the evaluation of $(\tilde{\Psi},{\cal{N}}(f^u))$ from (\ref{eq:projnl}) and
involves the Fourier components of $f^u$
\begin{equation}
f^u_l=\left[A\psi(v)\,\delta_{l,k} +
A^\ast\psi(v)^\ast\,\delta_{l,-k}\right] + H_l(v,A,A^\ast).
\label{eq:fufc}
\end{equation}
Combining (\ref{eq:projnl}) with (\ref{eq:hdef}) and (\ref{eq:fufc}) yields
\begin{eqnarray}
(\tilde{\Psi},{\cal{N}}(f^u))&=&\frac{-iA\sigma}{k}
\left\{<\partial_v\tilde{\psi},\kappa\cdot (h_0-h_2)>+\frac{\Gamma_{{2}}}{2}
<\partial_v\tilde{\psi},\kappa\cdot {\psi}^\ast> \right.\label{eq:peval}\\
&& +\sigma\left[\rule{0in}{0.2in}\Gamma_{{1}}
<\partial_v\tilde{\psi},\kappa\cdot  h_0> -\Gamma^\ast_{{1}}
<\partial_v\tilde{\psi},\kappa\cdot  h_2>+\frac{\Gamma_{{2}}}{2}
<\partial_v\tilde{\psi},\kappa\cdot  h^\ast_1>\right.\nonumber\\
&&\hspace{1.5in}\left.-\frac{\Gamma^\ast_{{2}}}{2}
<\partial_v\tilde{\psi},\kappa\cdot  h_3>\right]\nonumber\\
&&\left.+\sum_{l=3}^{\infty}\;\frac{\sigma^{l-2}}{l}[\Gamma_{{l}}
<\partial_v\tilde{\psi},\kappa\cdot  h^\ast_{l-1}>-\sigma\Gamma^\ast_{{l}}
<\partial_v\tilde{\psi},\kappa\cdot  h_{l+1}>]\right\}.\nonumber
\end{eqnarray}
In this expression, sum over the velocity integrals of $h^{(s)}_{m}$ is denoted
by
\begin{equation}
\Gamma_{{m}}(\sigma)\equiv\sum_s
\int^\infty_{-\infty}\,dv\,h^{(s)}_{m}(v,\sigma).\label{eq:Gdef}
\end{equation}
Now comparing (\ref{eqn:pdef}) and (\ref{eq:peval}) provides the desired
expression for $p$
\begin{eqnarray}
p(\sigma)&=&\lambda-\frac{i\sigma}{k}\left\{
<\partial_v\tilde{\psi},\kappa\cdot (h_0-h_2)>+\frac{\Gamma_{{2}}}{2}
<\partial_v\tilde{\psi},\kappa\cdot {\psi}^\ast> \right.\label{eq:ph}\\
&& +\sigma\left[\rule{0in}{0.2in}\Gamma_{{1}}
<\partial_v\tilde{\psi},\kappa\cdot  h_0> -\Gamma^\ast_{{1}}
<\partial_v\tilde{\psi},\kappa\cdot  h_2>+\frac{\Gamma_{{2}}}{2}
<\partial_v\tilde{\psi},\kappa\cdot  h^\ast_1>-\frac{\Gamma^\ast_{{2}}}{2}
<\partial_v\tilde{\psi},\kappa\cdot  h_3>\right]\nonumber\\
&&\left.+\sum_{l=3}^{\infty}\;\frac{\sigma^{l-2}}{l}[\Gamma_{{l}}
<\partial_v\tilde{\psi},\kappa\cdot  h^\ast_{l-1}>-\sigma\Gamma^\ast_{{l}}
<\partial_v\tilde{\psi},\kappa\cdot  h_{l+1}>]\right\}.\nonumber
\end{eqnarray}

This expression for $p$ involves the functions $h_m(v,\sigma)$ which define
$H$. The equations that determine $h_m(v,\sigma)$ follow from the general
equation for $H$ in (\ref{eq:Heqn}). In presenting these relations
we shall use ${\cal P}_\perp$,
the orthogonal projection defined by the unstable mode.
Let  ${\cal P}$ denote the projection operator onto $\psi(v)$
of a multi-component function $g(v)$,
\begin{equation}
({\cal P}g)(v)\equiv<\tilde{\psi},g>\psi(v),\label{eq:Pdef}
\end{equation}
with the orthogonal projection ${\cal P}_\perp\equiv I-{\cal P}$ given by
\begin{eqnarray}
({\cal P}_\perp g)(v)&=&g(v)-\psi(v)
<\tilde{\psi},g>.\label{eq:Pk}
\end{eqnarray}

With the previous expression for $(\tilde{\Psi},{\cal{N}}(f^u))$ in
(\ref{eq:peval}) and the notation in (\ref{eq:hdef}) for the Fourier components
of $H$, the components of (\ref{eq:Heqn}) take the form:
\begin{eqnarray}
\lefteqn{(p+p^\ast)\left[h_0+\sigma\frac{\partial h_0}{\partial
{\sigma}}\right]=}\hspace{0.5in}\label{eq:hfc0f}\\
&&\frac{i}{k}\,\frac{\partial}{\partial v}
\kappa\cdot \left\{\left[{\psi}^\ast+\sigma( {h_1}^\ast -\psi
\Gamma^\ast_{{1}})+\sigma^2\,
{h_1}^\ast \Gamma_{{1}}
+\sum^{\infty}_{l=2}\, \frac{\sigma^{l-1}}{l}
 {h_l}^\ast \Gamma_{{l}} \right]-cc\right\}\nonumber
\end{eqnarray}
\begin{eqnarray}
\lefteqn{\left[
(2p+p^\ast)h_1-(L_kh_1)
+(p+p^\ast)\sigma\frac{\partial h_1}{\partial
{\sigma}}\right]=}\hspace{1.0in}\label{eq:hfc1f}\\
&&\frac{i}{k}\,{\cal P}_\perp \frac{\partial}{\partial v}
\kappa\cdot \left\{\rule{0in}{0.3in}h_0-h_2+\frac{1}{2}{\psi}^\ast\,
\Gamma_{{2}}
+\sigma\left[h_0 \Gamma_{{1}} -
h_2\Gamma^\ast_{{1}}+\frac{1}{2}{h_1}^\ast\Gamma_{{2}}
-\frac{1}{2}h_3\Gamma^\ast_{{2}} \right]\right.\nonumber\\
&&\hspace{1.0in}+\sum^{\infty}_{l=3}\left.\frac{\sigma^{l-2}}{l}
\left[{h^{\ast}_{l-1}} \Gamma_{{l}}-\sigma h_{l+1} \Gamma^\ast_{{l}}\right]
\rule{0in}{0.3in}\right\}\nonumber
\end{eqnarray}
\begin{eqnarray}
\lefteqn{\left[
2p\,h_2-L_{2k}h_2
+(p+p^\ast)\sigma\frac{\partial h_2}{\partial
{\sigma}}\right]=}\hspace{1.0in}\label{eq:hfc2f}\\
&&\frac{i}{k}\,\frac{\partial}{\partial v}
\kappa\cdot\left\{\rule{0in}{0.3in}\psi +
\sigma\left[h_1+\psi \Gamma_{{1}} -h_3
+\frac{1}{2} h_0 \Gamma_{{2}} +\frac{1}{3} {\psi}^\ast \Gamma_{{3}}
\right]\right.
\nonumber\\
&&\hspace{1.25in}
\sigma^2\left[h_1\Gamma_{{1}}- h_3 \Gamma^\ast_{{1}}
-\frac{1}{2} h_4 \Gamma^\ast_{{2}} +\frac{1}{3} {h_1}^\ast \Gamma_{{3}}
\right]\nonumber\\
&&\hspace{1.0in}\left.
-\frac{\sigma^3}{3} h_5 \Gamma^\ast_{{3}}+
\sum^{\infty}_{l=4}\frac{\sigma^{l-2}}{l}\left[{h^{\ast}_{l-2}} \Gamma_{{l}}
-\sigma^2 h_{l+2} \Gamma^\ast_{{l}}\right]\rule{0in}{0.3in}\right\}\nonumber
\end{eqnarray}
and
\begin{eqnarray}
\lefteqn{\left[
mp\,h_m- L_{mk}h_m
+(p+p^\ast)\sigma\frac{\partial h_m}{\partial
{\sigma}}\right]=}\hspace{1.0in}\label{eq:hfckf}\\
&&\frac{i}{k}\,\frac{\partial}{\partial v}
\kappa\cdot\left\{\rule{0in}{0.3in}h_{m-1}+\frac{\psi}{m-1}\Gamma_{{m-1}}
+\sum^{m-2}_{l=2}\, \frac{h_{m-l}}{l}\Gamma_{{l}}\right.
\nonumber\\
&&\hspace{0.75in}+\sigma\left[h_{m-1}\Gamma_{{1}}-h_{m+1}
+\frac{h_1}{m-1}\Gamma_{{m-1}}
+\frac{h_0}{m}\Gamma_{{m}} +\frac{{\psi}^\ast}{m+1}\Gamma_{{m+1}}\right]
\nonumber\\
&&\hspace{0.75in}+\sigma^2\left[-h_{m+1}\Gamma^\ast_{{1}}
+\frac{{h_1}^\ast}{m+1}\Gamma_{{m+1}}\right]
+\sum^{\infty}_{l=m+2}\frac{\sigma^{l-m}}{l}{h^{\ast}_{l-m}}
\Gamma_{{l}} \nonumber\\
&&\hspace{1.5in}\left.-\sum^{\infty}_{l=2}\frac{\sigma^{l}}{l}h_{m+l}
\Gamma^\ast_{{l}}\rule{0in}{0.3in}\right\}\nonumber
\end{eqnarray}
for $m=0, 1, 2$, and $m>2$, respectively.

Together with (\ref{eq:ph}) these component equations determine the functions
$p(\sigma)$ and $\{h_m(\sigma)\}_{m=0}^{\infty}$; however they cannot
be solved except using the expansions introduced in the next section.
{}From a practical standpoint, we have achieved a reduction of the
problem to the analysis of functions of a {\em single} real variable, i.e.
$\sigma$.  In the study of the amplitude equation (\ref{eq:aeqn})
this reduction represents a very useful simplification.
%____________________________________________________________________

\section{Expansions, recursion relations, and pinching
singularities}\label{sec:pinch}
%______________________________________________________________________
The amplitude equation on the unstable manifold,
\begin{equation}
\dot{A}=A\;p(\sigma),\label{eq:modeeqn}
\end{equation}
is analyzed by expressing $p(\sigma)$, $h_m(v,\sigma)$,  and
$\Gamma_{{m}}(\sigma)$ as power series in $\sigma$:
\begin{equation}
p(\sigma)=\sum^{\infty}_{j=0}\,p_j\,\sigma^j
\hspace{0.5in}
h_m(v,\sigma)=\sum^{\infty}_{j=0}\,h_{m,j}(v)\,\sigma^j
\hspace{0.5in}
\Gamma_{{m}}(\sigma)=\sum^{\infty}_{j=0}\,\Gamma_{{m},{j}}
\,\sigma^j\label{eq:series}
\end{equation}
where $\Gamma_{{m},{j}}=\int dv \sum_s \,h^{(s)}_{m,j}(v)$.

The coefficients $p_j$ and $h_{m,j}(v)$ are calculated by recursively
solving (\ref{eq:ph}) and (\ref{eq:hfc0f}) - (\ref{eq:hfckf}). A recursion
relation
for $p_j$ follows from (\ref{eq:ph}) by inserting the expansions above and
solving at each order in $\sigma$; this calculation yields
\begin{eqnarray}
p_0&=&\lambda\label{eq:p0}\\
p_j&=&\frac{i}{k}[{\cal A}_{j}+{\cal B}_{j}]\;\;\;
\mbox{for}\;\;j\geq 1\label{eq:pj}
\end{eqnarray}
where
\begin{eqnarray}
{\cal A}_j&=&-<\partial_v\tilde{\psi},\kappa\cdot  (h_{0,j-1}-h_{2,j-1})>
-\frac{1}{2}<\partial_v\tilde{\psi}, \kappa\cdot
{\psi}^\ast>\Gamma_{{2},{j-1}}
\label{eq:ap}\\
&&\hspace{0.5 in}
-\sum^{j-2}_{l=0}\left[<\partial_v\tilde{\psi},\kappa\cdot
h_{0,j-l-2}>\Gamma_{{1},{l}}
-<\partial_v\tilde{\psi},\kappa\cdot  h_{2,j-l-2}>\Gamma^\ast_{{1},{l}}
\right]\nonumber\\
{\cal B}_j&=&-\sum^{j-2}_{l=0}
\left\{\frac{1}{2}<\partial_v\tilde{\psi}, \kappa\cdot  {h^{\ast}_{1,j-l-2}}>
\Gamma_{{2},{l}}\right. \label{eq:bp}\\
&&\hspace{0.5 in} +\left.
\sum^{l}_{m=0}\left[\frac{\Gamma_{{j-l+1},{m}}}{j-l+1}
<\partial_v\tilde{\psi}, \kappa\cdot  {h^{\ast}_{j-l,l-m}}>
-\frac{\Gamma^\ast_{{j-l},{m}}}{j-l}
<\partial_v\tilde{\psi},\kappa\cdot  h_{j-l+1,l-m}>\right]\right\} \nonumber
\end{eqnarray}
Here and below, a summation is understood to be omitted if the lower limit
exceeds the upper limit. This organization of the terms in (\ref{eq:ap}) and
(\ref{eq:bp}) will turn out to distinguish to different singular behaviors:
${\cal A}_j\sim \gamma^{-(5j-1)}$ and ${\cal B}_j\sim\gamma^{-(5j-2)}$.

The corresponding relations for $h_{m,j}(v)$ are determined similarly by
applying the expansions (\ref{eq:series}) to the general equations
(\ref{eq:hfc0f}) - (\ref{eq:hfckf}); the resulting equations are solved for
$h_{m,j}(v)$.
Our results are conveniently stated in terms of
the resolvent operator
$R_l(w)\equiv(w-L_l)^{-1}$ and certain auxiliary functions $I_{m,j}(v)$ whose
detailed expressions are provided below.
For $m=0$,  $h_{0,j}$ is simply
\begin{equation}
h_{0,j}(v)=\frac{I_{0,j}(v)}{(1+j)(\lambda+\lambda^\ast)},
\label{eq:h0lfcn}
\end{equation}
and the coefficients $h_{m,j}$ for $m>0$ require the resolvent
\begin{equation}
h_{m,j}(v)=R_{mk}(w_{m,j})\,I_{m,j}\label{eq:coeffhkl}
\end{equation}
with the complex numbers $w_{m,j}$ defined by
\begin{equation}
w_{m,j}\equiv(j+\delta_{m,1})(\lambda+\lambda^\ast)+m\lambda=
2(j+\delta_{m,1})\,\gamma+m\lambda.\label{eq:wkl}
\end{equation}

A general expression for $R_l(w)$ follows from (\ref{eq:linop}) by solving
$(w-L_l)f=g$ for $f$.\cite{cra1,cra2}  Here both $g(v)=(g^{(s_1)}(v),
g^{(s_2)}(v),\ldots)$
and $f$ are multi-component fields, and  $R_l(w)$
acts by
\begin{equation}
R_l(w)\,g=\left(
\begin{array}{c}(R_l(w)\,g)^{(s_1)}(v)\\(R_l(w)\,g)^{(s_2)}(v)\\\vdots
\end{array}\right)
\end{equation}
where
\begin{equation}
(R_l(w)\,g)^{(s)}(v)=\frac{1}{il(v-iw/l)}
\left[g^{(s)}(v)-\frac{\kappa^{(s)}\eta^{(s)}_l}{\Lambda_{l}(iw/l)}
\sum_{s'}\int^{\infty}_{-\infty}
\,dv'\,\frac{g^{(s')}(v')}{v'-iw/l}\right].\label{eq:resol}
\end{equation}

Note the following significant feature: the arguments $w_{m,j}$ of the
resolvent in (\ref{eq:coeffhkl})
determine {\em poles} of $h_{m,j}(v)$ located at $v=z_{m,j}$ where
$z_{m,j}\equiv{iw_{m,j}}/{mk}$. For $m\geq1$, these poles always fall in the
upper half-plane above the phase velocity $v_p$:
\begin{equation}
z_{m,j}= z_0+\frac{i\gamma d_{m,j}}{k}=v_p
+\frac{i\gamma(1+d_{m,j})}{k} \label{eq:respole}
\end{equation}
where $d_{m,j}\equiv 2(j+\delta_{m,1})/m$.

%-----------------------------------------------------------
\subsection{Recursion relations for $I_{m,j}$}
%-----------------------------------------------------------
The auxiliary functions $I_{0,j}(v)$ are defined by
\begin{eqnarray}
I_{0,0}(v)&=& \frac{i}{k}\,\frac{\partial }{\partial v}\kappa\cdot
({\psi}^\ast-\psi)
\label{eq:I00}
\end{eqnarray}
$j=0$, and for $j\geq1$
\begin{eqnarray}
I_{0,j}(v)&=& -\sum^{j-1}_{n=0}\,(1+n)(p_{j-n}+p_{j-n}^\ast)
\,h_{0,n}(v)\label{eq:I0l}\\
&&+\frac{i}{k}\,\frac{\partial}{\partial v}
\kappa\cdot\left\{\left[{h^{\ast}_{1,j-1}} -
\psi \Gamma^\ast_{{1},{j-1}}
+\sum^{j-2}_{n=0}\,{h^\ast_{1,n}}\Gamma_{{1},{j-n-2}}
\right.\right.\nonumber\\
&&\hspace{1.0in}\left.\left.
+\sum^{j-1}_{n=0}\,\sum^{n}_{l=0}\,\frac{{h^{\ast}_{j-n+1,l}}}{j-n+1}
\Gamma_{{j-n+1},{n-l}}\right]-cc\right\}.\nonumber
\end{eqnarray}

For $m=1$, the functions $I_{1,j}(v)$ are given by
\begin{eqnarray}
\lefteqn{I_{1,j}(v)=
-\sum^{j-1}_{n=0}\left[(2+n)p_{j-n}+(1+n)p_{j-n}^\ast\right]h_{1,n}}
\hspace{0.5in}\label{eq:I1l}
\\
&&+ \frac{i}{k}\,{\cal P}_\perp \frac{\partial}{\partial v}
\kappa\cdot\left\{\rule{0in}{0.3in}
h_{0,j}-h_{2,j}+ \frac{1}{2}{\psi}^\ast\Gamma_{{2},{j}}\right.\nonumber\\
&&\hspace{1.0in}
+\sum^{j-1}_{n=0}\left[
h_{0,n}\Gamma_{{1},{j-n-1}}-
h_{2,n}\Gamma^\ast_{{1},{j-n-1}}+\frac{1}{2}{{h^\ast_{1,n}}}
\Gamma_{{2},{j-n-1}}
\right.
\nonumber\\
&&\hspace{1.25in}\left.-\frac{1}{2}h_{3,n}\Gamma^\ast_{{2},{j-n-1}}
+\sum^{n}_{m=0}
\left(\frac{{h^{\ast}_{j-n+1,m}}}{j-n+2} \Gamma_{{j-n+2},{n-m}}
\right)\right]\nonumber\\
&&\hspace{1.0in}\left.
-\sum^{j-2}_{n=0}\sum^{n}_{m=0}
\left[\frac{h_{j-n+2,m}}{j-n+1} \Gamma^\ast_{{j-n+1},{n-m}}\right]
\rule{0in}{0.3in}\right\}.\nonumber
\end{eqnarray}
The notation ${\cal P}_\perp g(v)$ was defined in (\ref{eq:Pk}):
 ${\cal P}_\perp g(v)=g(v)-\psi(v)
<\tilde{\psi},g>$.

For $m=2$ the functions $I_{2,j}(v)$ are given by
\begin{eqnarray}
I_{2,0}(v)&=& \frac{i}{k}\,\frac{\partial}{\partial
v}\kappa\cdot\psi\label{eq:I20}
\end{eqnarray}
for $j=0$, and for $j\geq1$
\begin{eqnarray}
\lefteqn{I_{2,j}(v)=-\sum^{j-1}_{n=0}
\left[(2+n)p_{j-n}+np_{j-n}^\ast\right]\,h_{2,n}}
\hspace{0.5in}\label{eq:I2l}\\
&&+\frac{i}{k}\,\frac{\partial}{\partial v}\kappa\cdot\left\{
h_{1,j-1}+\psi \Gamma_{{1},{j-1}} -h_{3,j-1}
+\frac{1}{3} {\psi}^\ast \Gamma_{{3},{j-1}}+ \frac{1}{2} \sum^{j-1}_{n=0}
h_{0,n} \Gamma_{{2},{j-n-1}} \right.
\nonumber\\
&&\hspace{0.5in}+\sum^{j-2}_{n=0}\left[h_{1,n}\Gamma_{{1},{j-n-2}}-
h_{3,n}\Gamma^\ast_{{1},{j-n-2}}
-\frac{1}{2} h_{4,n} \Gamma^\ast_{{2},{j-n-2}}+\frac{1}{3} h^{\ast}_{1,n}
\Gamma_{{3},{j-n-2}}\right.
\nonumber\\
&&\hspace{2.5in}
+\sum^{n}_{m=0}\left.\frac{h^{\ast}_{j-n,m}}{j-n+2}
\Gamma_{{j-n+2},{n-m}}\right]
\nonumber\\
&&\hspace{0.5in}\left.
-\sum^{j-3}_{n=0}
\frac{h_{5,n} }{3} \Gamma^\ast_{{3},{j-n-3}}
-\sum^{j-4}_{n=0}\sum^{n}_{m=0}\frac{h_{j-n+2,m}}{j-n}
\Gamma^\ast_{{j-n},{n-m}}\right\}.\nonumber
\end{eqnarray}

For $m>2$, the functions $I_{m,j}(v)$ are given by
\begin{eqnarray}
I_{m,j}(v)&=&
-\sum^{j-1}_{n=0}\,[(m+n)p_{j-n}+n\,p_{j-n}^\ast]\,h_{m,n}(v)
\label{eq:Ikl}\\
&&\hspace{0.25in}+\frac{i}{k}\,\frac{\partial}{\partial v}
\kappa\cdot\left\{
h_{m-1,j}+\frac{\psi}{m-1}\Gamma_{{m-1},{j}}
+\sum^{m-2}_{l=2}\,\sum^{j}_{n=0}\, \frac{h_{m-l,n}}{l}\Gamma_{{l},{j-n}}
\right.\nonumber\\
&&\hspace{1.0in}-h_{m+1,j-1}+\frac{{\psi}^\ast}{m+1}
\Gamma_{{m+1},{j-1}}.
\nonumber\\
&&\hspace{1.0in}
+\sum^{j-1}_{n=0}\,
\left[{h_{m-1,n}}\Gamma_{{1},{j-n-1}}+\frac{h_{1,n}}{m-1}
\Gamma_{{m-1},{j-n-1}}+\frac{h_{0,n}}{m}\Gamma_{{m},{j-n-1}}\right]
\nonumber\\
&&\hspace{1.0in}+
\sum^{j-2}_{n=0}\,\left[-h_{m+1,n}\Gamma^\ast_{{1},{j-n-2}}
+\frac{h^{\ast}_{1,n}}{m+1}\Gamma_{{m+1},{j-n-2}}\right.\nonumber\\
&&\hspace{1.0in}\left.\left.
+\sum^{n}_{l=0}\,\left(\frac{h^{\ast}_{j-n,l}}{j+m-n}\Gamma_{{j+m-n},{n-l}}
-\frac{h_{m+j-n,l}}{j-n}\Gamma^\ast_{{j-n},{n-l}}\right)\right]\right\};
\nonumber
\end{eqnarray}
in this last expression, if a subscript is negative the term is understood to
be omitted, e.g. for $j=0$, $h_{m+1,j-1}$ is omitted.
%---------------------------------------------------------------
\subsection{Useful identities}
%---------------------------------------------------------------
The relations (\ref{eq:pj}), (\ref{eq:h0lfcn}) - (\ref{eq:coeffhkl}), and
(\ref{eq:I00}) - (\ref{eq:Ikl}) can be
applied systematically to calculate $p_j$ and $h_{m,j}$ to any order. The
leading coefficient $p_0$ is determined by linear
theory  (\ref{eq:p0}) and from the linear eigenfunction $\psi$ one can also
calculate $h_{0,0}$ and $h_{2,0}$, c.f. (\ref{eq:I00}) and (\ref{eq:I20}),
respectively. These two coefficients then suffice to calculate $p_1$ from
(\ref{eq:pj}).
{}From $\{p_1, h_{0,0}, h_{2,0}\}$, the coefficients $h_{1,0}$ and $h_{3,0}$
can be determined and then $h_{0,1}$ and $h_{2,1}$. This provides the input
to calculate $p_2$, and from this point on the structure of the calculation to
all orders falls into the simple pattern summarized in Table I.

For the remainder of the paper, our analysis of the
expansion (\ref{eq:series}) is facilitated by several identities which we
summarize here.
These relations allow $\Gamma_{{m},{j}}$, $h_{m,j}$, and certain
important integrals to be obtained very simply from the auxiliary functions
$I_{m,j}$.

If $m=0$ we note that (\ref{eq:h0lfcn}) and (\ref{eq:I00}) - (\ref{eq:I0l})
imply $\Gamma_{{0},{j}}=0$. For $m>0$, by substituting (\ref{eq:coeffhkl}) into
$\Gamma_{{m},{j}}=\int dv\, \sum_s h^{(s)}_{m,j}$ and rearranging, we obtain
\begin{eqnarray}
\Gamma_{{m},{j}}&=&\frac{-i/mk}{\Lambda_{mk}(z_{m,j})}\sum_s
\int^\infty_{-\infty}\,dv\,\frac{I^{(s)}_{m,j}(v)}{v-z_{m,j}}
\hspace{0.2in}(m>0).
\label{eq:gmkj}
\end{eqnarray}
 With (\ref{eq:gmkj}) for $\Gamma_{{m},{j}}$, the general form for $h_{m,j}$ in
(\ref{eq:coeffhkl}) can be re-expressed as
\begin{equation}
h_{m,j}(v)=\left(\frac{-i}{mk}\right)\frac{I_{m,j}(v)}{v-z_{m,j}}
-\left(\frac{\kappa\cdot\eta_{mk}}{v-z_{m,j}}\right)\Gamma_{{m},{j}}
\hspace{0.2in}(m>0).\label{eq:hmj}
\end{equation}
In this way $\Gamma_{{m},{j}}$ and $h_{m,j}$ are written explicitly in terms of
$I_{m,j}$.

The coefficients $p_j$ depend on integrals of the form
$<\partial_v\tilde{\psi},\kappa\cdot h_{m,j}>$, and it is helpful to express
these integrals directly in terms of $I_{m,j}$ also.
{}From (\ref{eq:h0lfcn}) we obtain the identity
\begin{equation}
<\partial_v\tilde{\psi},\kappa\cdot h_{0,j}>=
\frac{1}{2\gamma(1+j)\Lambda'_{k}(z_0)}
\int_{-\infty}^{\infty}\,dv
\frac{\sum_s\kappa^{(s)}I^{(s)}_{0,j}}{(v-z_{0})^2};\label{eq:h0j}
\end{equation}
similarly, by integrating (\ref{eq:hmj}) we find
\begin{eqnarray}
<\partial_v\tilde{\psi},\kappa\cdot h_{m,j}>&=&
\left(\frac{-i}{mk\Lambda'_{k}(z_0)}\right)
\int_{-\infty}^{\infty}\,dv
\frac{\sum_s\kappa^{(s)}I^{(s)}_{m,j}}{(v-z_{0})^2(v-z_{m,j})}\nonumber\\
&&\hspace{0.5in}-\left(\frac{\Gamma_{{m},{j}}}{\Lambda'_{k}(z_0)}\right)
\int_{-\infty}^{\infty}\,dv
\frac{\sum_s(\kappa^{(s)})^2\eta^{(s)}_{mk}}{(v-z_{0})^2(v-z_{m,j})}.
\label{eq:hmjint}
\end{eqnarray}

%---------------------------------------------------------------
\subsection{Analysis of the cubic coefficient}\label{subsec:cubic}
%---------------------------------------------------------------
It is instructive at this point to evaluate and examine the cubic coefficient
$p_1$ in detail. This coefficient illustrates the occurrence of pinching
singularities and the procedures needed to
evaluate such singular integrals.

{}From (\ref{eq:pj}) we have
\begin{equation}
p_1=-\frac{i}{k}\left[<\partial_v\tilde{\psi},\kappa\cdot(h_{0,0}-h_{2,0})>
+\frac{\Gamma_{2,0}}{2}<\partial_v\tilde{\psi},\kappa\cdot\psi^\ast>\right]
\label{eq:p1coeff}
\end{equation}
so $h_{0,0}(v)$ , $h_{2,0}(v)$ and $\Gamma_{2,0}$ are needed to calculate
$p_1$.
{}From (\ref{eq:h0lfcn}) and (\ref{eq:I00}) we have
\begin{equation}
h_{0,0}(v)= -\frac{1}{k^2}\frac{\partial}{\partial v}
\left[\frac{\kappa^2\cdot\eta_k}{(v-z_0)(v-z^\ast_0)} \right]
\label{eq:h00}
\end{equation}
where $\kappa^2$ denotes the square of the mass matrix.
{}From (\ref{eq:specfcnid}), (\ref{eq:I20}) and (\ref{eq:gmkj}) we obtain
\begin{equation}
\Gamma_{2,0}=-\frac{2}{3k^{2}}\int_{-\infty}^{\infty}\,dv \frac{\sum_s
(\kappa^2\cdot\eta_k)^{(s)}}{(v-z_0)^3},
\label{eq:gam20}
\end{equation}
and  inserting (\ref{eq:I20}) into  (\ref{eq:hmj}) yields
\begin{equation}
 h_{2,0}(v)=\frac{1}{2k^2}\left(\frac{\kappa\cdot\partial_v \psi}{v-z_0}\right)
 -\frac{\Gamma_{2,0}}{4} \left(\frac{ \kappa\cdot\eta_k}{v-z_0}\right).
\label{eq:h20}
\end{equation}
Here we have made the substitution $\eta_{2k}=\eta_k/4$.

We are concerned
with the form of the various integrals in $p_1$ in the asymptotic regime of
small growth rate $\gamma\rightarrow0^+$.  The behavior of these integrals is
determined by any complex singularities of the integrand that approach the real
velocity axis
as $\gamma\rightarrow0^+$. We shall always assume that whatever complex
analytic
structure characterizes $F_0$ there are never any singularities that move
to the real axis in this limit and therefore the singularities in $\eta_k$ can
be regarded as irrelevant in analyzing a given integral.

The singularities of interest are associated with the
explicit poles in the denominator of a given integrand.
For example,  $\Gamma_{2,0}$
has a third-order pole at $v=z_0$ which approaches the real axis from above
as $\gamma\rightarrow0^+$. In this case, there are no corresponding poles
in the lower half-plane, and consequently no pinching singularity develops;
therefore $\Gamma_{2,0}$ has a finite limit as the growth rate goes to zero.
Similarly, we see by inspection that $h_{2,0}$ has a third-order pole at
$v=z_0$
and consequently the integrand of
$<\partial_v\tilde{\psi},\kappa\cdot h_{2,0}>$ has a fifth-order pole at
$v=z_0$. However, again there is no pinching singularity and this integral also
has a finite limit as $\gamma\rightarrow0^+$.

The remaining integrals, $<\partial_v\tilde{\psi},\kappa\cdot\psi^\ast>$ and
$<\partial_v\tilde{\psi},\kappa\cdot h_{0,0}>$, are more interesting. The
simplest example is $<\partial_v\tilde{\psi},\kappa\cdot\psi^\ast>$ whose
integrand has a double pole at $v=z_0$ and a simple pole at $v=z_0^\ast$:
\begin{equation}
<\partial_v\tilde{\psi},\kappa\cdot\psi^\ast>=-\frac{1}{\Lambda'_k(z_0)}
\sum_s\int^\infty_{-\infty}\frac{dv\,(\kappa^2\cdot\eta_k)^{(s)}}
{(v-z_0)^2(v-z_0^\ast)}.\label{eq:csing1}
\end{equation}
As $\gamma\rightarrow0^+$, these poles approach the real axis from above and
below;
this creates a pinching singularity and the integral has a singular limit.
We analyze this integral in two steps that can be applied to all
such singular integrals: a partial fraction expansion extracts the singularity
and
the dispersion relation is applied to eliminate integrals over the electron
distribution function. The motivation for the second step is to incorporate the
one relation between the different distribution functions and to expose the
effect of treating the ions as mobile.

The partial fraction expansion of the integrand yields
\begin{equation}
\frac{1}
{(v-z_0)^2(v-z_0^\ast)}=\left(\frac{ik}{2\gamma}\right)^2
\left[\frac{1}{v-z_0^\ast}-\frac{1}{v-z_0}\right]-
\left(\frac{ik}{2\gamma}\right)\frac{1}{(v-z_0)^2}
\end{equation}
so that our integral becomes
\begin{eqnarray}
\sum_s\int^\infty_{-\infty}\frac{dv\,(\kappa^2\cdot\eta_k)^{(s)}}
{(v-z_0)^2(v-z_0^\ast)}&=&
\left(\frac{ik}{2\gamma}\right)^2
\left[\sum_s\int^\infty_{-\infty}\frac{dv\,(\kappa^2\cdot\eta_k)^{(s)}}
{v-z_0^\ast}-\sum_s\int^\infty_{-\infty}\frac{dv\,(\kappa^2\cdot\eta_k)^{(s)}}
{v-z_0}\right]\label{eq:pfexp}\\
&&-\left(\frac{ik}{2\gamma}\right)
\sum_s\int^\infty_{-\infty}\frac{dv\,(\kappa^2\cdot\eta_k)^{(s)}}
{(v-z_0)^2}.\nonumber
\end{eqnarray}
On the right hand side of this expression all the integrals are free of
pinching singularities and therefore have finite limits
as $\gamma\rightarrow0^+$. The singularities of the original integral are
now explicitly shown in the factors $\gamma^{-2}$ and $\gamma^{-1}$.

The dispersion relation for the unstable mode $\Lambda_k(z_0)=0$ implies the
identities
\begin{eqnarray}
\int^\infty_{-\infty}\,dv\,\frac{\eta_k^{(e)}(v)}{v-z_0}&=&
1+\int^\infty_{-\infty}\,dv\,
\frac{\sum_s'\kappa^{(s)}\eta_k^{(s)}(v)}{v-z_0}\label{eq:dr}\\
\int^\infty_{-\infty}\,dv\,\frac{\eta_k^{(e)}(v)}{(v-z_0)^2}&=&
-\Lambda_k'(z_0)-\int^\infty_{-\infty}\,dv\,
\frac{\sum_s'\kappa^{(s)}\eta_k^{(s)}(v)}{(v-z_0)^2}
\label{eq:drp}
\end{eqnarray}
where we have used $\kappa^{(e)}=-1$ and the primed sum indicates that the
electron term is omitted. Substitution of (\ref{eq:dr}) - (\ref{eq:drp}) for
the electron integrals in (\ref{eq:pfexp}) yields
\begin{eqnarray}
\sum_s\int^\infty_{-\infty}\frac{dv\,(\kappa^2\cdot\eta_k)^{(s)}}
{(v-z_0)^2(v-z_0^\ast)}&=&
\left(\frac{ik}{2\gamma}\right)^2
\left[-2i{\sum_s}'\kappa^{(s)}(1+\kappa^{(s)})\;{\rm Im}
\int^\infty_{-\infty}\frac{dv\,\eta_k^{(s)}}
{v-z_0}\right]\nonumber\\
&&+\left(\frac{ik}{2\gamma}\right)\left[\Lambda_k'(z_0)
+{\sum_s}'\kappa^{(s)}(1-\kappa^{(s)})
\int^\infty_{-\infty}\frac{dv\,\eta_k^{(s)}}
{(v-z_0)^2}\right].\label{eq:close}
\end{eqnarray}
Combining this with (\ref{eq:csing1}) we conclude that
$<\partial_v\tilde{\psi},\kappa\cdot\psi^\ast>$ typically diverges like
$\gamma^{-2}$ in the small growth rate limit. There are exceptions however;
for example, in the limit of fixed ions, $\kappa^{(s)}\rightarrow0$ for $s\neq
e$, then we are left with
\begin{equation}
\sum_s\int^\infty_{-\infty}\frac{dv\,(\kappa^2\cdot\eta_k)^{(s)}}
{(v-z_0)^2(v-z_0^\ast)}\rightarrow \left(\frac{ik}{2\gamma}\right)
\Lambda_k'(z_0),
\end{equation}
and the divergence is merely $\gamma^{-1}$.

We treat the remaining integral $<\partial_v\tilde{\psi},\kappa\cdot h_{0,0}>$
in the same fashion. This integrand now has a fourth order pole
at $v=z_0$ and a simple pole at $v=z_0^\ast$:
\begin{equation}
<\partial_v\tilde{\psi},\kappa\cdot h_{0,0}>=-\frac{2}{k^2\Lambda'_k(z_0)}
\sum_s\int^\infty_{-\infty}\frac{dv\,(\kappa^3\cdot\eta_k)^{(s)}}
{(v-z_0)^4(v-z_0^\ast)}.\label{eq:csing2}
\end{equation}
Again we make a partial fraction expansion
\begin{eqnarray}
\frac{1}{(v-z_0)^4\,(v-z_0^\ast)}&=&
\left(\frac{ik}{2\gamma}\right)^4\left[\frac{1}{v-z_0^\ast}-\frac{1}{v-z_0}
\right]\\
&&-\left(\frac{ik}{2\gamma}\right)^3\frac{1}{(v-z_0)^2}
-\left(\frac{ik}{2\gamma}\right)^2\frac{1}{(v-z_0)^3}
-\left(\frac{ik}{2\gamma}\right)\frac{1}{(v-z_0)^4},\nonumber
\end{eqnarray}
and eliminate the electron integrals to find
\begin{eqnarray}
 \sum_s\int^\infty_{-\infty}\frac{dv\,(\kappa^3\cdot\eta_k)^{(s)}}
{(v-z_0)^4(v-z_0^\ast)} &=&
\left(\frac{ik}{2\gamma}\right)^4
\left[+2i{\sum_s}'\kappa^{(s)}(1-{\kappa^{(s)}}^2)\;{\rm Im}
\int^\infty_{-\infty}\frac{dv\,\eta_k^{(s)}}
{v-z_0}\right]\label{eq:cubint1}\\
&&+\left(\frac{ik}{2\gamma}\right)^3\left[-\Lambda_k'(z_0)
+{\sum_s}'\kappa^{(s)}(1-{\kappa^{(s)}}^2)
\int^\infty_{-\infty}\frac{dv\,\eta_k^{(s)}}
{(v-z_0)^2}\right]\nonumber\\
&&+\left(\frac{ik}{2\gamma}\right)^2
\left[-\frac{\Lambda_k''(z_0)}{2}
+{\sum_s}'\kappa^{(s)}(1-{\kappa^{(s)}}^2)
\int^\infty_{-\infty}\frac{dv\,\eta_k^{(s)}}
{(v-z_0)^3}\right]\nonumber\\
&&+\left(\frac{ik}{2\gamma}\right)
\left[-\frac{\Lambda_k'''(z_0)}{6}
+{\sum_s}'\kappa^{(s)}(1-{\kappa^{(s)}}^2)
\int^\infty_{-\infty}\frac{dv\,\eta_k^{(s)}}
{(v-z_0)^4}\right].\nonumber
\end{eqnarray}
Thus for $<\partial_v\tilde{\psi},\kappa\cdot h_{0,0}>$ the typical divergence
is $\gamma^{-4}$.

Combining these results with $p_1$ in (\ref{eq:p1coeff}), we obtain
the asymptotic form of the cubic coefficient as $\gamma\rightarrow0^+$
\begin{eqnarray}
p_1&=&\frac{1}{\gamma^4}
\left[c_1(\gamma)-\gamma\,d_1(\gamma) + {\cal O}(\gamma^2)\right]
\label{eq:p1asymp}
\end{eqnarray}
where $c_1$ and $d_1$ are nonsingular functions of $\gamma$ defined by
\begin{eqnarray}
c_1(\gamma)&=&-\frac{k}{4\Lambda_k'(z_0)}
{\sum_s}'{\kappa^{(s)}(1-{\kappa^{(s)}}^2)}\;{\rm
Im}\left(\int^\infty_{-\infty}\,dv\frac{\eta_k^{(s)}}{v-z_0}\right)
\label{eq:a1}\\
d_1(\gamma)&=&\frac{1}{4}-
\frac{1}{4\Lambda_k'(z_0)}{\sum_s}'\kappa^{(s)}(1-{\kappa^{(s)}}^2)
\int^\infty_{-\infty}\,dv\frac{\eta_k^{(s)}}{(v-z_0)^2};\label{eq:b1}
\end{eqnarray}
this result was briefly presented without derivation in an
earlier paper.\cite{jdcaj} When the leading term is
nonzero at $\gamma=0$ (i.e. $c_1(0)\neq0$), then $p_1$ diverges like
$\gamma^{-4}$ as a result
of the divergence in $<\partial_v\tilde{\psi},\kappa\cdot h_{0,0}>$.

The significance of this result is best understood in light of our earlier
discussion of scaling the amplitude $A(t)$ to obtain an amplitude equation
free of singular features at small growth rates. With our expression for
$p_1$, the amplitude equation for the unstable mode is
\begin{eqnarray}
\dot{A}&=&A\left[\lambda+\frac{[c_1(0)+ {\cal O}(\gamma)]}{\gamma^4} |A|^2 +
p_2(\gamma) |A|^4+\cdots\right]
\end{eqnarray}
which we rewrite using polar variables $A=\rho e^{-i\theta}$ as
\begin{eqnarray}
\dot{\rho}&=& \gamma\; \rho+ \frac{1}{\gamma^4}[{\rm Re}(c_1(0)) +{\cal
O}(\gamma)]\;
\rho^3 + {\rm Re}(p_2)\; \rho^5+\cdots \\
\dot{\theta}&=&\omega - \frac{1}{\gamma^4}[{\rm Im}(c_1(0))+ {\cal
O}(\gamma)]\;\rho^2 - {\rm Im}(p_2)\; \rho^4+\cdots.
\end{eqnarray}
Now we introduce a scaled amplitude $\rho(t)=\gamma^\beta r(\gamma t)$
following (\ref{eq:newvar}) and anticipate that
$\delta=1$ is required to remove the
singular effect of the linear term:
\begin{eqnarray}
\frac{dr}{d\tau}&=& r +\frac{ \gamma^{2\beta-1}}{\gamma^4}[{\rm Re}(c_1(0))
+{\cal O}(\gamma)]\;
r^3 + \gamma^{4\beta-1}{\rm Re}(p_2)\; r^5+\cdots\label{eq:recscale1} \\
\dot{\theta}&=&\omega - \frac{\gamma^{2\beta}}{\gamma^4}[{\rm Im}(c_1(0))+
{\cal O}(\gamma)]\;r^2 -\gamma^{4\beta} {\rm Im}(p_2)\; r^4+\cdots.
\end{eqnarray}
The exponent $\beta$ should be chosen so that the nonlinear coefficients have
finite limits as $\gamma\rightarrow0^+$. The most stringent
requirements on $\beta$ arise from the coefficients in the $dr/d\tau$ equation,
and the cubic term requires
$2\beta\geq5$ to absorb the $\gamma^{-4}$ singularity.
Setting $\beta=5/2$ suffices to cancel the singularities in $p_1$
and allow the cubic nonlinearity to formally balance with the linear term at
small $\gamma$:
\begin{eqnarray}
\frac{dr}{d\tau}&=& r + [{\rm Re}(c_1(0)) +{\cal O}(\gamma)]\;
r^3 + \gamma^{9}{\rm Re}(p_2)\; r^5+\cdots \label{eq:rescale}\\
\dot{\theta}&=&\omega - \gamma\; [{\rm Im}(c_1(0))+ {\cal O}(\gamma)]\;r^2
-\gamma^{10} {\rm Im}(p_2)\; r^4+\cdots.
\end{eqnarray}

This choice for $\beta$ assumes  that ${\rm Re}(c_1(0))\neq0$
and that the singularities at fifth (and higher)
order are also removed by the same scaling that absorbs
the cubic singularity. Typically the condition ${\rm Re}(c_1(0))\neq0$ is
satisfied, but there are interesting exceptions and they are discussed  next.
The adequacy of the $\beta=5/2$ scaling in controlling higher order
singularities is studied in section \ref{sec:all orders}.

%------------------------------------------------------------
\subsection{Special Cases: $c_1(0)=0$}\label{subsec:special}
%------------------------------------------------------------

Evaluating  $c_1(0)$ as $\gamma\rightarrow0^+$ yields
\begin{equation}
c_1(0)=-\frac{\pi\,k}{4\Lambda_k'(v_p+i0)}
{\sum_s}'{\kappa^{(s)}(1-{\kappa^{(s)}}^2)}\;\eta_k^{(s)}(v_p(0))
\label{eq:gzero}
\end{equation}
where $v_p(0)$ denotes the linear phase velocity at zero growth rate. Since the
first factor
$(\pi\,k/4\Lambda_k')$ cannot vanish, the special case $c_1(0)=0$ will only
arise if the
species sum (which excludes the electrons) vanishes. By carefully adjusting the
parameters of
a given model, one can presumably arrange for exact cancellations between
different terms in
this sum, but there is little motivation to study such artificial situations.

However, there are
three natural circumstances in which $c_1(0)=0$ because every term in the sum
vanishes independently, and the  cubic coefficient is clearly less singular:
\begin{enumerate}
\renewcommand{\theenumi}{\alph{enumi}}
\item Infinitely massive ions: $\kappa^{(s)}=0$ for all $s\neq e$.
\item Zero slope for the resonant ions: $\eta_k^{(s)}(v_p(0))=0$ for all
$s\neq e$.
\item An electron-positron plasma: ${\kappa^{(p)}}^2=1$ for
positrons.
\end{enumerate}
In all three cases, the cubic coefficient has the asymptotic form
\begin{eqnarray}
p_1&=&-\frac{1}{\gamma^3}
\left[d_1(0) + {\cal O}(\gamma)\right]
\label{eq:p1asympsc}
\end{eqnarray}
with $d_1(0)$ given by the full formula (\ref{eq:b1}) for the second case and
$d_1(0)=1/4$ in the first and third cases.

In the rescaled equation (\ref{eq:recscale1}), this
weakened singularity requires only that $2\beta\geq4$, and setting $\beta=2$
yields
\begin{eqnarray}
\frac{dr}{d\tau}&=& r + [{\rm Re}(d_1(0)) +{\cal O}(\gamma)]\;
r^3 + \gamma^{7}{\rm Re}(p_2)\; r^5+\cdots \label{eq:rdot}\\
\dot{\theta}&=&\omega - [{\rm Im}(d_1(0))+ {\cal O}(\gamma)]\;r^2 -\gamma^{8}
{\rm Im}(p_2)\; r^4+\cdots. \label{eq:tdot}
\end{eqnarray}
In (I), we have shown that the exponent $\beta=2$ suffices to absorb the
singularities in (\ref{eq:rdot}) - (\ref{eq:tdot}) to {\em all} orders for
a plasma with fixed ions; this exponent corresponds to the
relatively familiar trapping scaling for the electric field of an unstable
mode. In Section \ref{sec:special}, we provide a new proof of this
result and also establish the same conclusion for the case (b) where each ion
distribution
is flat at the phase velocity: $\eta_k^{(s)}(v_p)=0$ for all $s\neq e$.
This will occur, for example,  if the ions are sufficiently cold and do not
populate the
resonant region in velocity space. On the other hand, when there is an exact
reflection symmetry
$F_0(v)=F_0(-v)$, we can have an unstable mode with $v_p=0$ and
$\eta_k^{(s)}(0)=0$ holds even though the ion distribution function is non-zero
at $v=0$.
This latter situation arises  for instabilities of reflection-symmetric
equilibria due
to real eigenvalues, e.g. a reflection-symmetric two-stream instability.

Our understanding of the third example, an electron-positron plasma,
is less complete. In Section \ref{subsec:ep} we obtain by explicit calculation
${\rm Re}(p_2)\sim\gamma^{-8}$ so the rescaled fifth order term in
(\ref{eq:rdot}) remains singular unless $\beta\geq9/4$. However, the
singularity
structure of the full expansion is more complicated and we have not
determined the effects of singularities at higher order.

%_________________________________________________________________________

\section{Singularity structure of the expansion}\label{sec:all orders}
%_______________________________________________________________________

Our goal is a systematic analysis of the singularities of $p_j$ to all orders
in the amplitude equation (\ref{eq:modeeqn}).
The detailed calculation of $p_j$ rapidly becomes prohibitively
laborious, but the recursion relations (\ref{eq:pj}) - (\ref{eq:bp})
determining the higher order
coefficients in terms of lower order quantities (cf. Table I) can be analyzed
to determine the properties of $p_j$.
 This study requires an accurate
estimate of the pinching singularities in the integrals found in
$p_j$. For this purpose we introduce an ``index'' which allows the divergence
of
a given integral to be assessed by a simple counting procedure. As in our
discussion of the cubic coefficient, we assume the analytic singularities of
$F_0$ do not affect the divergence behavior;
the precise assumption on $F_0$ is given in (\ref{eq:F0ok}) below.
%------------------------------------------------------------
\subsection{Definition of the index}\label{subsec:indexdef}
%------------------------------------------------------------
At every order in the expansion, the pinching singularities have a common
structure. For $n>0$, define
\begin{equation}
D_n(\alpha,v)\equiv\frac{1}{(v-\alpha_1)(v-\alpha_2)\cdots(v-\alpha_n)}
\label{eq:Ddef}
\end{equation}
where  $\alpha\equiv(\alpha_1,\ldots,\alpha_n)$ and also define
$D_0(\alpha,v)\equiv1$. Evaluating $p_j$ for $j\geq1$ involves integrands of
the form
\begin{equation}
{\cal G}(v)=D_m(\beta,v)^\ast\,D_n(\alpha,v)\,
\sum_s\;(\kappa^{(s)})^{m'}\;\frac{\partial^q\;\eta_l^{(s)}}{\partial v^q}
\label{eq:indatom}
\end{equation}
with $m+n\geq1$, $m'\geq1$ and $q\geq0$.
The poles in ${\cal G}(v)$ may be written as
\begin{eqnarray}
\alpha_j&=&z_0+i\gamma\nu_j/k\hspace{0.5in}j=1,\ldots,n\label{eq:poles1}\\
\beta^\ast_j&=&
z_0^\ast-i\gamma\zeta_j/k\hspace{0.5in}j=1,\ldots,m;\label{eq:poles2}
\end{eqnarray}
since they always lie along the vertical line ${\rm  Re}(v)=v_p(\gamma)$.
The specific pole locations are given by the numbers
$\nu_j\geq0$ and $\zeta_j\geq0$ which vary depending on the integral.
In all cases, these numbers are independent
of $F_0$; in particular $\nu_j$ and $\zeta_j$ are independent of
$\gamma$.

The complex-analytic singularities of $F_0$ are assumed to be
unimportant for the amplitude
expansions in the following sense. For any $n\geq1$ and  $q\geq0$,
we assume that
\begin{equation}
\lim_{\gamma\rightarrow0^+}\left|\int^\infty_{-\infty}\,dv\,
D_n(\alpha,v)\frac{\partial^q\;\eta_l^{(s)}}{\partial v^q}\right|<\infty
\label{eq:F0ok}
\end{equation}
holds for each species and arbitrary choices of $\nu_j$ in
(\ref{eq:poles1}). For example,
it is sufficient for each distribution function $F_0^{(s)}(v)$  to be analytic
on
a neighborhood of $v_p(0)$, the linear phase velocity at zero growth rate; in
this case
the limit in (\ref{eq:F0ok}) can be evaluated by the Plemej formulas for Cauchy
integrals.\cite{musk}

The {\em index} of ${\cal G}(v)$ in (\ref{eq:indatom}) is defined by
\begin{equation}
{\mbox{\rm Ind }} [{\cal G}]\equiv m+n+q-1,\label{eq:inddef}
\end{equation}
and  ${\mbox{\rm Ind }} [{\cal G}]\geq0$
since we assume $m+n\geq1$. For example, for $\psi$ in (\ref{eq:lefcn}), the
integrand of $\Gamma_{2,0}$ in (\ref{eq:gam20}) and integrand of
$<\partial_v\tilde{\psi},\kappa\cdot\psi^\ast>$ in (\ref{eq:csing1}),
we obtain the indices
\begin{equation}
{\mbox{\rm Ind }}\left[\sum_s \kappa^{(s)}\psi^{(s)}\right]=0\hspace{0.3in}
{\mbox{\rm Ind }} \left[\frac{\sum_s
(\kappa^2\cdot\eta_k)^{(s)}}{(v-z_0)^3}\right]=2\hspace{0.3in}
{\mbox{\rm Ind }} \left[\frac{\sum_s\,(\kappa^2\cdot\eta_k)^{(s)}}
{(v-z_0)^2(v-z_0^\ast)}\right]=2.
\label{eq:indexex1}
\end{equation}

If $mn=0$, our assumption (\ref{eq:F0ok}) ensures that $\int dv{\cal G}(v)$
is non-singular, but when $mn\neq0$,
then
as ${\gamma\rightarrow0^+}$, the integral of ${\cal G}$ develops a
pinching singularity at the phase velocity $v_p(0)$. The index of ${\cal G}$
estimates the worst possible  divergence of $\int dv\,{\cal G}$ due to this
singularity.

\begin{lemma}\label{lem:index} For ${\cal G}(v)$ in \mbox{\rm
(\ref{eq:indatom})} with $mn\neq0$, the integral of ${\cal G}$
satisfies
\begin{equation}
\lim_{{\gamma\rightarrow0^+}}\; \gamma^{J}\;\left|
\int^\infty_{-\infty}\,dv\,{\cal G}(v)\right|<\infty\label{eq:sing0}
\end{equation}
with $J={\mbox{\rm Ind }} [{\cal G}]$. However, if
\begin{equation}
{\sum_s}^\prime\kappa^{(s)}[(-1)^{m'}+(\kappa^{(s)})^{m'-1}]\;
\eta^{(s)}_{k}(v_p(0))=0,
\label{eq:special}
\end{equation}
then the integral of ${\cal G}$ is less singular and \mbox{\rm
(\ref{eq:sing0})} holds with $J={\mbox{\rm Ind }} [{\cal G}]-1$. In \mbox{\rm
(\ref{eq:special})} the prime on the summation indicates that the sum omits the
electron species.
\end{lemma}
\noindent {\em {\bf Proof}.}
\begin{quote}
 By integrating by parts $q$ times we can reduce to the case with $q=0$. With
$q=0$, consider the simplest possiblity
$m+n=2$, and make a partial fraction expansion of the integrand
\begin{eqnarray}
\int^\infty_{-\infty}\,dv\,
\frac{\sum_s\;(\kappa^{(s)})^{m'}\;\eta_l^{(s)}}
{(v-\alpha_1)(v-\beta_1^\ast)}&=&\frac{(-ik/\gamma)}{(2+\nu_1+\zeta_1)}
\left[\int^\infty_{-\infty}\,dv\,
\frac{\sum_s\;(\kappa^{(s)})^{m'}\;\eta_l^{(s)}}
{(v-\alpha_1)}\right.\\
&&\left.\hspace{1.0in}-
\int^\infty_{-\infty}\,dv\,
\frac{\sum_s\;(\kappa^{(s)})^{m'}\;\eta_l^{(s)}}
{(v-\beta_1^\ast)}\right].\nonumber
\end{eqnarray}
The integrals on the right are non-singular with limits at $\gamma=0$ given by
the Plemej
formula
\begin{eqnarray}
\lefteqn{\lim_{\gamma\rightarrow0^+}
\left[\int^\infty_{-\infty}\,dv\,
\frac{\sum_s\;(\kappa^{(s)})^{m'}\;\eta_l^{(s)}}
{(v-\alpha_1)}-
\int^\infty_{-\infty}\,dv\,
\frac{\sum_s\;(\kappa^{(s)})^{m'}\;\eta_l^{(s)}}
{(v-\beta_1^\ast)}\right]=}\label{eq:limit}\\
&&\hspace{3.0in}
2\pi\,i\sum_s\;(\kappa^{(s)})^{m'}\;\eta_l^{(s)}(v_p(0)).\nonumber
\end{eqnarray}
{}From the  limit $\gamma\rightarrow0^+$ of the dispersion relation
$\Lambda_{k}(z_0)=0$, we obtain
$\sum_s\;(\kappa^{(s)})\;\eta_k^{(s)}(v_p(0))=0$ for the imaginary part. Since
$\kappa^{(e)}=-1$, this can be rewritten as
$\eta_k^{(e)}=\sum'_s\;(\kappa^{(s)})\;\eta_k^{(s)}(v_p(0))$, and used to
eliminate $\eta_k^{(e)}$ from (\ref{eq:limit}):
\begin{eqnarray}
\lefteqn{\lim_{\gamma\rightarrow0^+}
\left[\int^\infty_{-\infty}\,dv\,
\frac{\sum_s\;(\kappa^{(s)})^{m'}\;\eta_l^{(s)}}
{(v-\alpha_1)}-
\int^\infty_{-\infty}\,dv\,
\frac{\sum_s\;(\kappa^{(s)})^{m'}\;\eta_l^{(s)}}
{(v-\beta_1^\ast)}\right]=}\label{eq:limit2}\\
&&\hspace{1.5in}
2\pi\,i\left(\frac{k^2}{l^2}\right){\sum_s}'\;\kappa^{(s)}
[(-1)^{m'}+(\kappa^{(s)})^{m'-1}]\;\eta_k^{(s)}(v_p(0)).\nonumber
\end{eqnarray}
This proves the lemma for $m+n=2$. It is important that the limit
in (\ref{eq:limit2}) is independent of the parameters
$\nu_1$ and $\zeta_1$ that locate the poles $\alpha_1$ and $\beta_1$ for
$\gamma>0$.

For integrals with  $m+n>2$, by expanding the integrand in partial
fractions, they
can be re-expressed in terms of integrals with $m+n-1$ multiplied by
a factor of $\gamma^{-1}$.
A simple induction argument then establishes (\ref{eq:sing0}) for general
$mn\neq0$. (Recall that if $mn=0$ then the integrals are nonsingular.)

The general significance of the condition in (\ref{eq:special}) becomes clear
from the
form of the partial fraction expansion. For example, consider
the expansion for $(m,n)=(2,1)$:
\begin{eqnarray}
\lefteqn{\int^\infty_{-\infty}\,dv\,
\frac{\sum_s\;(\kappa^{(s)})^{m'}\;\eta_l^{(s)}}
{(v-\alpha_1)(v-\alpha_2)(v-\beta_1^\ast)}=
\frac{(-ik/\gamma)}{(2+\nu_1+\zeta_1)}
\int^\infty_{-\infty}\,dv\,
\frac{\sum_s\;(\kappa^{(s)})^{m'}\;\eta_l^{(s)}}
{(v-\alpha_1)(v-\alpha_2)}}\\
&&\nonumber\\
&&\hspace{0.5in}-\frac{(-ik/\gamma)^2}{(2+\nu_1+\zeta_1)(2+\nu_2+\zeta_1)}
\left[\int^\infty_{-\infty}\,dv\,
\frac{\sum_s\;(\kappa^{(s)})^{m'}\;\eta_l^{(s)}}
{(v-\alpha_2)}-
\int^\infty_{-\infty}\,dv\,
\frac{\sum_s\;(\kappa^{(s)})^{m'}\;\eta_l^{(s)}}
{(v-\beta_1^\ast)}\right]\nonumber
\end{eqnarray}
The second term is the dominant singularity $\gamma^{-2}$ and at $\gamma=0$
the bracketed integrals yield the same factor obtained previously in
(\ref{eq:limit2}). When this factor is zero, the singularity of the
$(m,n)=(2,1)$ integral is reduced from 2 to 1.

In the partial fraction expansion of the general case, the most singular
terms that diverge according to (\ref{eq:sing0}) will always be proportional
to the difference of two non-singular integrals:
\begin{equation}
\left[\int^\infty_{-\infty}\,dv\,
\frac{\sum_s\;(\kappa^{(s)})^{m'}\;\eta_l^{(s)}}
{(v-\alpha_j)}-
\int^\infty_{-\infty}\,dv\,
\frac{\sum_s\;(\kappa^{(s)})^{m'}\;\eta_l^{(s)}}
{(v-\beta_{j'}^\ast)}\right]
\end{equation}
whose limit at $\gamma=0$ is given by (\ref{eq:limit2}). Thus when this
limit is zero, the singularity of the term must drop by 1. This proves
the second part of the lemma.

{\bf $\Box$}\end{quote}

The exceptional situation defined by the condition in (\ref{eq:special}) is
a generalization of the feature noted previously in our analysis of the cubic
coefficient. Setting $m'=3$ in (\ref{eq:special}) yields  $c_1(0)=0$
from (\ref{eq:gzero}); the special circumstance that corresponded to a less
singular integral at third order.

Our application of Lemma \ref{lem:index} to the recursion relations for
$I_{m,j}$ requires generalization of the index to allow for sums of functions
with well-defined indices and
products of ${\cal G}$ with singular functions of $\gamma$. In each case, the
generalized index is defined so that Eq. (\ref{eq:sing0}) remains true, i.e.
the index for the composite function gives the {\em maximal} possible
divergence of its integral.

First, if ${\cal G}_1(v)$
and ${\cal G}_2(v)$ have indices satisfying ${\mbox{\rm Ind }} [{\cal
G}_1]\geq{\mbox{\rm Ind }} [{\cal G}_2]$ then
we define the index of the sum to be the larger index:
\begin{equation}
{\mbox{\rm Ind }} [{\cal G}_1+{\cal G}_2]\equiv{\mbox{\rm Ind }} [{\cal
G}_1].\label{eq:sum}
\end{equation}
Clearly, ${\mbox{\rm Ind }} [{\cal G}_1]$ gives
the maximal possible divergence of
$\int\,dv\,({\cal G}_1+{\cal G}_2)$. Secondly,
if $q(\gamma)$ is a function of $\gamma$
with the asymptotic behavior $q(\gamma)\sim\gamma^{-\nu}$
as ${\gamma\rightarrow0^+}$,
then we define the index of $q(\gamma)\,{\cal G}(v)$ to be
\begin{equation}
{\mbox{\rm Ind }} [q\,{\cal G}]\equiv {\mbox{\rm Ind }} [{\cal
G}]+\nu.\label{eq:prod}
\end{equation}
Estimates of the form (\ref{eq:sing0}) still hold for $q{\cal G}$ with
$J={\mbox{\rm Ind }} [q\,{\cal G}]$.
This completes the definition of the index.

By applying (\ref{eq:sum}) and (\ref{eq:prod}), the
indices of $I_{m,j}$ and $h_{m,j}$ may be determined.
For example, from (\ref{eq:I00}) and (\ref{eq:I20}) we have the indices
\begin{equation}
{\mbox{\rm Ind }} \left[\sum_s \kappa^{(s)}I^{(s)}_{0,0}\right]=1\hspace{0.5in}
{\mbox{\rm Ind }} \left[\sum_s
\kappa^{(s)}I^{(s)}_{2,0}\right]=1,\label{eq:indexex2}
\end{equation}
and from (\ref{eq:h00}) and (\ref{eq:h20})
\begin{equation}
{\mbox{\rm Ind }} \left[\sum_s \kappa^{(s)}h^{(s)}_{0,0}\right]=2\hspace{0.5in}
{\mbox{\rm Ind }} \left[\sum_s
\kappa^{(s)}h^{(s)}_{2,0}\right]=2.\label{eq:indexex3}
\end{equation}
The index for $h^{(s)}_{2,0}$ is correct provided $\Gamma_{2,0}$
does not diverge more strongly than $\gamma^{-2}$. We can confirm
this immediately from
the index in (\ref{eq:indexex1}); in fact, $\Gamma_{2,0}$ is non-singular, as
noted already in Section \ref{subsec:cubic}.
In section \ref{subsec:examples}, these indices are applied to determine the
singularity of $p_1$.
We stress again that
when applied to a composite function ${\cal G}(v)$ the estimate in
(\ref{eq:sing0}) does not
necessarily determine the true singularity, but only an upper bound on
the possible divergence of the integral.

There are several immediate consequences of the index definition worth stating.
First, complex conjugation doesn't alter the index,
${\mbox{\rm Ind }} [{\cal G}]={\mbox{\rm Ind }} [{\cal G}^\ast];$
secondly, if $G(v)$ has a well-defined index,
then dividing $G$ by $(v-\alpha)$ or $(v-\beta^\ast)$
simply increases the index of $G$ by one:
\begin{equation}
{\mbox{\rm Ind }} [{\cal G}/(v-\alpha)]=
{\mbox{\rm Ind }} [{\cal G}/(v-\beta^\ast)]=
{\mbox{\rm Ind }} [{\cal G}]+1.\label{eq:divide}
\end{equation}
Here $\alpha, \beta$ are defined as in (\ref{eq:poles1}) - (\ref{eq:poles2})
and thereby increase the index of every term in $G$ by one; this implies
(\ref{eq:divide}). Finally, from (\ref{eq:indatom}), differentiating ${\cal G}$
also raises the index by one,
${\mbox{\rm Ind }} \left[\partial_v{\cal G}\right]=1+
{\mbox{\rm Ind }}\left[{\cal G}\right]$.

%-----------------------------------------------------------
\subsection{Analysis of $p_1$ using the index}\label{subsec:examples}
%-----------------------------------------------------------

The above relations along with recursion relations determine the indices of
$I_{m,j}$ and
$h_{m,j}$ to all orders. This information permits an estimate of the
singularities of the coefficients $p_j$ for $j\geq1$. In practice, this
must be done recursively following the organization of the amplitude expansions
in Table I. The procedure
is illustrated
here for the first level of Table I; more general results
on the index to all orders are provided in Section \ref{subsec:main} and
Section \ref{sec:special}.

Table I begins with $\psi$, $h_{0,0}$ and $h_{2,0}$ whose indices were
evaluated in (\ref{eq:indexex1}) and (\ref{eq:indexex3}).
{}From this information,  we can estimate the integrals required to evaluate
$p_1$ in (\ref{eq:p1coeff}). The integrand of
$<\partial_v\tilde{\psi},\kappa\cdot\psi^\ast>$ has index 2, and the integrands
of $<\partial_v\tilde{\psi},\kappa\cdot h_{0,0}>$
and $<\partial_v\tilde{\psi},\kappa\cdot h_{2,0}>$ have index 4.
In each case, the factor $\partial_v\tilde{\psi}^{(s)^\ast}$ increases the
index by two from the values obtained in (\ref{eq:indexex1}) and
(\ref{eq:indexex3}). Now Lemma \ref{lem:index} tells us that
$<\partial_v\tilde{\psi},\kappa\cdot\psi^\ast>$ cannot diverge more strongly
than $\gamma^{-2}$, and $<\partial_v\tilde{\psi},\kappa\cdot h_{0,0}>$ and
$<\partial_v\tilde{\psi},\kappa\cdot h_{2,0}>$
cannot diverge more strongly than $\gamma^{-4}$.

These conclusions imply that
the singularity
of $p_1$ cannot be worse than $\gamma^{-4}$; this was found in
Section \ref{subsec:cubic} by an  explicit calculation of
$p_1$. The explicit evaluation, of course, provided the important
additional  result
that the $\gamma^{-4}$ singularity actually occurs unless one
of the exceptional cases corresponding to $c_1(0)=0$ is considered. The
foregoing index analysis is easier but yields less complete information.

%-----------------------------------------------------------
\subsection{The Main Result}\label{subsec:main}
%-----------------------------------------------------------
The main result on the singularity structure of the amplitude expansions
can now be proved.

\begin{theorem}\label{thm:main} For $j\geq1$, the coefficients in the
expansion of the amplitude equation \mbox{\rm (\ref{eq:modeeqn})} satisfy
\begin{equation}
\lim_{{\gamma\rightarrow0^+}}\;
\gamma^{5j-1}\;\left|p_j\right|<\infty.\label{eq:pjasy}
\end{equation}
Let $J_{m,j}\equiv(2m+5j-3)+4\delta_{m,0}+5\delta_{m,1}$, then for $j\geq0$,
$m\geq0$ and $m'\geq0$, the indices of $I^{(s)}_{m,j}$ and
$h^{(s)}_{m,j}$ obey
\begin{eqnarray}
{\mbox{\rm Ind }} \left[\sum_s(\kappa^{(s)})^{m'}I^{(s)}_{m,j}\right]&\leq&
J_{m,j}.\label{eq:keyind}\\
{\mbox{\rm Ind }} \left[\sum_s(\kappa^{(s)})^{m'}h^{(s)}_{m,j}\right]&\leq&
J_{m,j}+1,\label{eq:hind}
\end{eqnarray}
and the integrals in
{\rm (\ref{eq:gmkj})} and {\rm (\ref{eq:hmjint})} satisfy
\begin{eqnarray}
\lim_{{\gamma\rightarrow0^+}}\; \gamma^{J_{m,j}+1}\; \left|\Gamma_{m,j}
\right|&<&\infty\label{eq:gmklasy}\\
\lim_{{\gamma\rightarrow0^+}}\;\gamma^{J_{m,j}+3-\delta_{m,1}}\;
\left|<\partial_v\tilde{\psi},\kappa\cdot h_{m,j}>\right|
&<&\infty.\label{eq:intest}
\end{eqnarray}
In addition, integrals over $I^{(s)}_{1,j}(v)$ satisfy
\begin{equation}
\lim_{{\gamma\rightarrow0^+}}\; \gamma^{J_{1,j}+n-1}\;
\left|\int^\infty_{-\infty}\;dv D_n(\alpha,v)
\sum_s (\kappa^{(s)})^{m'}\,I^{(s)}_{1,j}(v)
\right|<\infty\label{eq:m=1}
\end{equation}
for $n\geq1$ where $D_n(\alpha,v)$ is defined in {\rm (\ref{eq:Ddef})} with
upper half-plane poles $\alpha=(\alpha_1,
\alpha_2,\ldots,\alpha_n)$ of the form {\rm (\ref{eq:poles1})} but otherwise
arbitrary.
\end{theorem}

The relations in
(\ref{eq:keyind}) - (\ref{eq:m=1}) are of interest chiefly for their utility in
proving (\ref{eq:pjasy}).
We are able to prove (\ref{eq:keyind})-(\ref{eq:hind}) as upper
bounds, but we expect
them to hold as equalities in most cases.
For $m=1$, the bounds in (\ref{eq:gmklasy}) and (\ref{eq:intest}) give
stronger estimates on the integrals
than one would obtain simply from the index of the integrand calculated using
(\ref{eq:keyind})-(\ref{eq:hind}). Similarly, the bounds in
(\ref{eq:m=1}) are needed because they provide sharper estimates for the $m=1$
case
than are available using the index information
(\ref{eq:keyind})-(\ref{eq:hind}) alone.

Even though
$h_{1,j}$ and $h_{1,j}^\ast$ have the same index, it is very important to
recognize that the conclusion in (\ref{eq:intest})
may not hold for $<\partial_v\tilde{\psi},\kappa\cdot h_{1,j}^\ast>$.
The most singular terms in
$<\partial_v\tilde{\psi},\kappa\cdot h_{1,j}>$ are not in general the most
singular terms in $<\partial_v\tilde{\psi},\kappa\cdot h_{1,j}^\ast>$; for
$<\partial_v\tilde{\psi},\kappa\cdot h_{1,j}^\ast>$,
if we wish to avoid explicitly evaluating the integral, our
only means of estimating the divergence is by calculating the index of the
integrand from (\ref{eq:hind}).

\noindent {\em {\bf Proof} .}\begin{quote}
\begin{enumerate}

\item The proof is by induction using the organization of
Table I. In Section \ref{subsec:indexdef}-\ref{subsec:examples},
the relations in (\ref{eq:pjasy}) -
(\ref{eq:intest}) have been explicitly verified for $I_{0,0}$, $I_{2,0}$,
$h_{0,0}$, $h_{2,0}$, $\Gamma_{0,0}$, $\Gamma_{2,0}$, and $p_1$.
This proves the theorem  for the first level of
Table I, but does not establish the validity of (\ref{eq:m=1}) for
the initial case $j=0$ which enters in the second level of Table I.
By substituting the recursion relation (\ref{eq:I1l}) for $I_{1,0}$ into
(\ref{eq:m=1}) we obtain
\begin{eqnarray}
\lefteqn{ \gamma^{3+n}\; \int^\infty_{-\infty}\;dv\, D_n(\alpha,v)
\sum_s (\kappa^{(s)})^{m'}\,I^{(s)}_{1,0}(v)=}\\
&&\hspace{0.5in}\gamma^{3+n}\left(\frac{i}{k}\right)
\int^\infty_{-\infty}\;dv\,D_n(\alpha,v){\partial_v}
\left(\kappa\cdot[h_{0,0}-h_{2,0}+
\frac{1}{2}{\psi}^\ast\Gamma_{{2},{0}}]\right)\nonumber\\
&&\hspace{0.75in}+\gamma^{3+n}\left(\frac{i}{k}\right)
<{\partial_v}\tilde{\psi},
\kappa\cdot[h_{0,0}-h_{2,0}+ \frac{1}{2}{\psi}^\ast\Gamma_{{2},{0}}]>
\int^\infty_{-\infty}\;dv\,D_n(\alpha,v)\psi(v).\nonumber
\end{eqnarray}
On the right hand side, the index of the first integrand is $(n+3)$
so the factor of $\gamma^{n+3}$ ensures a finite limit. In the second term,
the integral over $D_n\,\psi$ is non-singular since all poles are in
the upper half-plane, but the coefficient $<{\partial_v}\tilde{\psi},
\kappa\cdot[h_{0,0}-h_{2,0}+ \frac{1}{2}{\psi}^\ast\Gamma_{{2},{0}}]>$
diverges like $\gamma^{-4}$ (cf. Section \ref{subsec:cubic}). Thus
for $n\geq1$ the second term will also have a finite limit; this verifies
(\ref{eq:m=1}) for $I_{1,0}$.

\item The relations (\ref{eq:pjasy}) - (\ref{eq:m=1}) are extended
by induction to all coefficients $I_{m,j}$, $h_{m,j}$, and $p_j$ using the
recursion relations (\ref{eq:pj}) - (\ref{eq:bp}) and
(\ref{eq:I00}) - (\ref{eq:Ikl}).  Assume that  (\ref{eq:pjasy}) -
(\ref{eq:m=1}) are valid down to some arbitrary level of Table I,
and consider what the recursion
relations imply for the coefficients evaluated at the next level. We use
$(m_1,j_1)$ to denote subscripts of coefficients such as $I_{m_1,j_1}$,
$h_{m_1,j_1}$ and $p_{j_1}$ that are to be evaluated from lower order
quantities assumed to satisfy (\ref{eq:pjasy}) -
(\ref{eq:m=1}).

The induction argument has three parts. First, we show that the properties
(\ref{eq:hind}) - (\ref{eq:intest}) must hold for $h_{m_1,j_1}$ and
$\Gamma_{m_1,j_1}$ if  $I_{m_1,j_1}$
satisfies two relations, namely
the index identity  (\ref{eq:keyind})
\begin{equation}
{\mbox{\rm Ind }} \left[\sum_s (\kappa^{(s)})^{m'}I^{(s)}_{m_1,j_1}\right]\leq
J_{m_1,j_1},\label{eq:keyind2}
\end{equation}
and the $m_1=1$ estimate (\ref{eq:m=1})
\begin{equation}
\lim_{{\gamma\rightarrow0^+}}\; \gamma^{J_{1,j_1}+n-1}\;
\left|\int^\infty_{-\infty}\;dv D_n(\alpha,v)
\sum_s (\kappa^{(s)})^{m'}\,I^{(s)}_{1,j_1}(v)
\right|<\infty.\label{eq:indstep}
\end{equation}

In the second part of the proof, we verify that (\ref{eq:keyind2}) -
(\ref{eq:indstep}) do hold for $I_{m_1,j_1}$. Finally, in the third part of the
proof, we verify that (\ref{eq:hind}) -  (\ref{eq:intest}) imply
(\ref{eq:pjasy})  for $p_{j_1}$. It is important to note from Table I
that the quantities $h_{m_1,j_1}$ depend on $p_j$ for $j<j_1$, but are
independent of $p_{j_1}$ so the reasoning is
not circular.

\item {\bf (Part 1)}: Suppose that (\ref{eq:keyind2}) - (\ref{eq:indstep})
hold,
then the three relations (\ref{eq:hind}) - (\ref{eq:intest}) are easily
obtained as follows. First, consider
(\ref{eq:gmklasy}) which holds trivially for $m_1=0$ since $\Gamma_{0,j_1}=0$.
For
$m_1\geq2$, (\ref{eq:gmklasy}) follows from (\ref{eq:keyind2}) and the identity
(\ref{eq:gmkj}). For $m_1=1$, we have to allow for the fact that
$\Lambda_{k}(z_{1,j_1})\sim\gamma$ as ${\gamma\rightarrow0^+}$, and apply
(\ref{eq:indstep}) to estimate the integral in (\ref{eq:gmkj}).

Next (\ref{eq:hind}) follows
by applying (\ref{eq:keyind2}) and  (\ref{eq:gmklasy}) to the identity
(\ref{eq:hmj}). Each term in (\ref{eq:hmj}) separately has a maximum index
of $(J_{m_1,j_1}+1)$ so this must bound the index of the sum.

Finally consider (\ref{eq:intest}). For $m_1\neq1$, this relation follows from
(\ref{eq:hind}),  and for $m_1=1$ we evaluate (\ref{eq:intest}) using the
identity in  (\ref{eq:hmj}):
\begin{eqnarray}
\gamma^{J_{1,j_1}+2}\;
<\partial_v\tilde{\psi},\kappa\cdot h_{1,j_1}>&=&
-\gamma^{5j_1+6}\left(\frac{i}{k}\right)
<\partial_v\tilde{\psi},\kappa\cdot{I_{1,j_1}}/{(v-z_{1,j_1})}>\\
&&\hspace{0.2in}-\gamma^{5j_1+6}\Gamma_{{1},{j_1}}
<\partial_v\tilde{\psi},\kappa^2\cdot\eta_{k}/{(v-z_{1,j_1})}>\nonumber
\end{eqnarray}
The first term on the right is non-singular by virtue of (\ref{eq:indstep}),
and the second term is also non-singular by virtue of (\ref{eq:gmklasy}).

\item {\bf (Part 2)}: Thus the crux of the matter is to
verify (\ref{eq:keyind2}) and (\ref{eq:indstep}) from the recursion relations
for  $I^{(s)}_{m_1,j_1}$, i.e. from (\ref{eq:I00}) - (\ref{eq:Ikl}). In this
step, we assume (\ref{eq:pjasy}) - (\ref{eq:m=1}) apply to the quantities
appearing on the right hand side of
these recursion relations, then the properties in (\ref{eq:keyind2}) -
(\ref{eq:indstep}) are verified term by term.

{\em Verification of \mbox{\rm(\ref{eq:keyind2})} - the index of
$I_{m_1,j_1}$:}\\
 The index of each term on the right hand side of the
recursion relations for $I^{(s)}_{m_1,j_1}(v)$ can be evaluated by applying
(\ref{eq:pjasy}) -  (\ref{eq:intest}); this exercise shows that all of these
terms have an index less than or equal to $J_{m_1,j_1}$ which
establishes (\ref{eq:keyind2}) for $I^{(s)}_{m_1,j_1}$. A few examples from
(\ref{eq:I1l}), the recursion relation for $({m_1,j_1})=(1,j_1)$, illustrate
these index calculations; note that in these examples $J_{1,j_1}=5j_1+4$.
\begin{enumerate}

\item The terms in (\ref{eq:I1l}) that depend on $p_j$ have the form
$[(2+n)p_{j_1-n}+(1+n)p_{j_1-n}^\ast]h^{(s)}_{1,n}$; from (\ref{eq:hind}) we
have
\begin{equation}
{\mbox{\rm Ind }} \left[\sum_s (\kappa^{(s)})^{m'}\,h^{(s)}_{1,n}\right]
\leq J_{1,n}+1=5n+5,
\end{equation}
and from (\ref{eq:pjasy}) the singularity of $p_{j_1-n}$ is determined, hence
\begin{equation}
{\mbox{\rm Ind }} \left[p_{j_1-n}\sum_s
(\kappa^{(s)})^{m'}\,h^{(s)}_{1,n}\right]
\leq5(j_1-n)-1+(5n+5)=5j_1+4
\label{eq:index1}
\end{equation}
which is consistent with (\ref{eq:keyind2}) for $(m_1,j_1)=(1,j_1)$.

\item Next consider the term
${\cal P}_\perp\kappa\cdot\partial_v\,h_{0,j_1}$ in (\ref{eq:I1l}); from the
definition (\ref{eq:Pk}) we have that the index
of ${\cal P}_\perp\kappa\cdot\partial_v\,h_{0,j_1}$ is equal to the
largest index obtained from the two terms $\kappa\cdot\partial_v\,h_{0,j_1}$
and
$\psi(v) <\partial_v\tilde{\psi},\kappa\cdot h_{0,j_1}>$.
These terms have individual indices
\begin{eqnarray}
{\mbox{\rm Ind }} \left[\sum_s (\kappa^{(s)})^{m'}
\partial_v\,h^{(s)}_{0,j_1}\right]
&\leq&5j_1+3\label{eq:index2a}\\
{\mbox{\rm Ind }}\left[\sum_s (\kappa^{(s)})^{m'}
\psi^{(s)}(v)<\partial_v\tilde{\psi},\kappa\cdot h_{0,j_1}>
\right]&\leq&5j_1+4
\label{eq:index2b}
\end{eqnarray}
where the first
index (\ref{eq:index2a}) comes from (\ref{eq:hind}) and the second follows
from (\ref{eq:intest}).
Hence
\begin{equation}
{\mbox{\rm Ind }}\left[\sum_s (\kappa^{(s)})^{m'}
\left({\cal
P}_\perp\kappa\cdot\partial_v\,h_{0,j_1}\right)^{(s)}\right]\leq5j_1+4
\end{equation}
which is consistent with (\ref{eq:keyind2}) for $(m_1,j_1)=(1,j_1)$.

\item A more subtle example is the term
${\cal P}_\perp\,\kappa\cdot\partial_v\,{{h^\ast_{1,n}}}
\Gamma_{{2},{j_1-n-1}}$ in (\ref{eq:I1l})
which expands to
\begin{eqnarray}
{\cal P}_\perp\,\kappa\cdot\partial_v\,{{h^\ast_{1,n}}}
\Gamma_{{2},{j_1-n-1}}&=&
\kappa\cdot\partial_v\,{{h^\ast_{1,n}}}
\Gamma_{{2},{j_1-n-1}}\label{eq:indx3}\\
&&\hspace{0.3in}+\Gamma_{{2},{j_1-n-1}}\,\psi(v)
<\partial_v\tilde{\psi},\kappa\cdot{h^\ast_{1,n}}>.
\nonumber
\end{eqnarray}
Now the index of each term can be evaluated to give
\begin{eqnarray}
{\mbox{\rm Ind }} \left[\Gamma_{{2},{j_1-n-1}}\sum_s
(\kappa^{(s)})^{m'}\partial_v\,{{h^{(s)}_{1,n}}^\ast}
\right]&\leq&5j_1+2
\label{eq:index3a}\\
{\mbox{\rm Ind }} \left[\Gamma_{{2},{j_1-n-1}}\sum_s
(\kappa^{(s)})^{m'}\psi^{(s)}
<\partial_v\tilde{\psi},\kappa\cdot{h^\ast_{1,n}}>
\right]
\label{eq:index3b}
&\leq&5j_1+4;
\end{eqnarray}
thus
\begin{equation}
{\mbox{\rm Ind }} \left[\sum_s (\kappa^{(s)})^{m'}
\left({\cal P}_\perp\,\kappa\cdot\partial_v\,{{h^\ast_{1,n}}}\right)^{(s)}
\Gamma_{{2},{j_1-n-1}}\right]\leq5j_1+4
\end{equation}
which is consistent with (\ref{eq:keyind2}) for $(m_1,j_1)=(1,j_1)$. In this
last
example, (\ref{eq:index3a}) is obtained from (\ref{eq:hind}) and
(\ref{eq:gmklasy}); the second
result (\ref{eq:index3b}) differs from (\ref{eq:index2b}) since
(\ref{eq:intest}) cannot be applied and the singularity of
$<\partial_v\tilde{\psi},\kappa\cdot{h^\ast_{1,n}}>$ must be estimated from the
index in (\ref{eq:hind}).
\end{enumerate}
In a similar way the index of each term
appearing on the right in the recursion relations
(\ref{eq:I00}) - (\ref{eq:Ikl}) for $I_{m_1,j_1}(v)$
can be estimated with the final result in (\ref{eq:keyind2}).

\item {\em Verification of \mbox{\rm(\ref{eq:indstep})}:}\\
We integrate the recursion relation for $I_{1,j_1}(v)$ and
examine the
resulting expression to establish (\ref{eq:indstep}).
When (\ref{eq:I1l}), the recursion for $I_{1,j_1}$, is inserted into
(\ref{eq:indstep}) we obtain
\begin{eqnarray}
\gamma^{J_{1,j_1}+n-1}\lefteqn{\int^\infty_{-\infty}\;dv\,D_n(\alpha,v)\,
{\sum_s (\kappa^{(s)})^{m'}\,I^{(s)}_{1,j_1}(v)}=}\nonumber\\
&&\hspace{0.0in}-\gamma^{5j_1+n+3}
\sum^{j_1-1}_{l=0}
\left[(2+l)p_{j_1-l}+(1+l)p_{j_1-l}^\ast\right]
\int^\infty_{-\infty}\;dv\,D_n(\alpha,v)\,
{\sum_s (\kappa^{(s)})^{m'}\,h^{(s)}_{1,l}(v)}\nonumber
\\
&&\hspace{0.5in}+\gamma^{5j_1+n+3}\left(
\frac{i}{k}\right)\,\int^\infty_{-\infty}\;dv\,D_n(\alpha,v)\,
{\sum_s (\kappa^{(s)})^{m'+1}\,G^{(s)}(v)}\label{eq:I1lint}
\\
&&\hspace{0.5in}+ \gamma^{5j_1+n+3}\left(\frac{i}{k}\right)
<\partial_v\tilde{\psi},\kappa\cdot G>
\int^\infty_{-\infty}\;dv\,D_n(\alpha,v)\,
{\sum_s (\kappa^{(s)})^{m'}\,\psi^{(s)}(v)}\nonumber
\end{eqnarray}
where $G(v)=h_{0,j_1}-h_{2,j_1} +\cdots$ denotes the collection of terms
enclosed by brackets in (\ref{eq:I1l}). The relations
(\ref{eq:keyind}) -  (\ref{eq:intest}) hold for these terms by assumption,
and it is straightforward to verify that ${\mbox{\rm Ind }}
\left[G\right]\leq5j_1+2$; this immediately implies that
 the  terms in (\ref{eq:I1lint})
involving $G$ are finite as $\gamma\rightarrow0^+$.
For the remaining terms in (\ref{eq:I1lint}) we have
$p_{j_1-l}\sim\gamma^{-5(j_1-l)+1}$ from (\ref{eq:pjasy}), hence we
need to prove
\begin{equation}
\lim_{{\gamma\rightarrow0^+}}\;\gamma^{5l+4+n}
\left|\int^\infty_{-\infty}\;dv\,D_n(\alpha,v)\,
{\sum_s (\kappa^{(s)})^{m'}\,h^{(s)}_{1,l}(v)}\right|<\infty
\hspace{0.5in}(n\geq1)
\label{eq:crux2}
\end{equation}
to establish (\ref{eq:indstep}). The formula in (\ref{eq:hind})
bounds the index of the integrand in (\ref{eq:crux2}) by only $(5l+n+5)$
which is not sufficient. Instead, we evaluate (\ref{eq:crux2}) using the
identity (\ref{eq:hmj})
\begin{eqnarray}
\int^\infty_{-\infty}\;dv\,D_n(\alpha,v)\,
{\sum_s (\kappa^{(s)})^{m'}\,h^{(s)}_{1,l}(v)}&=&
-\left(\frac{i}{k}\right)
\int^\infty_{-\infty}\;dv\,D_n(\alpha,v)\,
\frac{\sum_s (\kappa^{(s)})^{m'}\,I^{(s)}_{1,l}(v)}{(v-z_{1,l})}\nonumber\\
&&\hspace{0.0in}
-\Gamma_{{1},{l}}
\int^\infty_{-\infty}\;dv\,D_n(\alpha,v)\,
\frac{\sum_s (\kappa^{(s)})^{m'+1}\,\eta^{(s)}_{k}}{(v-z_{1,l})}.
\label{eq:reduce}
\end{eqnarray}
The integral in the second term is non-singular and
$\Gamma_{{1},{l}}\sim\gamma^{-5l-5}$
 from (\ref{eq:gmklasy}), thus the
desired estimate  in  (\ref{eq:crux2}) holds for the second
term. It remains to verify
\begin{equation}
\lim_{{\gamma\rightarrow0^+}}\;\gamma^{5l+4+n}\;
\left|\int^\infty_{-\infty}\;dv\,D_n(\alpha,v)\,
\frac{\sum_s (\kappa^{(s)})^{m'}\,I^{(s)}_{1,l}(v)}{(v-z_{1,l})}\right|
<\infty\hspace{0.5in}(n\geq1)\label{eq:crux3}
\end{equation}
for the first term in (\ref{eq:reduce}); this follows from (\ref{eq:m=1}) with
$j\rightarrow l$ and $D_n\rightarrow D_n/(v-z_{1,l})$.

The proof of (\ref{eq:crux2}) establishes (\ref{eq:indstep}) for $I_{1,j_1}$.

\item {\bf (Part 3)}: {\em Verification of {\rm(\ref{eq:pjasy})}
for $p_{j_1}$}\\
It is straightforward to check that application of (\ref{eq:hind}) -
(\ref{eq:intest}) to
the right hand side of the recursion relations (\ref{eq:pj}) - (\ref{eq:bp})
yields the estimate in
(\ref{eq:pjasy}).
A few examples from (\ref{eq:ap}) - (\ref{eq:bp})
suffice to illustrate how this is carried out.
\begin{enumerate}
\item  Consider the term $<\partial_v\tilde{\psi},\kappa\cdot h_{0,j_1-1}>$ in
(\ref{eq:ap});
from (\ref{eq:intest}) the singularity of this integral is at most
\begin{equation}
\left|<\partial_v\tilde{\psi},\kappa\cdot h_{0,j_1-1}>\right|
\sim\left(\frac{1}{\gamma}\right)^{3+J_{0,j_1-1}}=
\left(\frac{1}{\gamma}\right)^{5j_1-1}
\label{eq:ex1}
\end{equation}
which is consistent with (\ref{eq:pjasy}).

\item Other terms will involve products of integrals; for example,
$\Gamma_{{2},{j_1-1}}<\partial_v\tilde{\psi},\kappa\cdot {\psi}^\ast>$ in
(\ref{eq:ap}).
Using (\ref{eq:pfexp}) and (\ref{eq:gmklasy}) we determine that
the singularity of this product is at most
\begin{equation}
\left|\Gamma_{{2},{j_1-1}}
<\partial_v\tilde{\psi},\kappa\cdot {\psi}^\ast>
\right|\sim\left(\frac{1}{\gamma}\right)^{1+J_{2,j_1-1}}
\left(\frac{1}{\gamma}\right)^2
=\left(\frac{1}{\gamma}\right)^{5j_1-1}
\label{eq:ex2}
\end{equation}
which again  is consistent with (\ref{eq:pjasy}).

\item A final example is the product
$\Gamma_{{2},{l}}<\partial_v\tilde{\psi},\kappa\cdot {h^{\ast}_{1,j_1-l-2}}>$
with $0\leq l\leq j_1-2$ from (\ref{eq:bp}); applying (\ref{eq:hind}) and
(\ref{eq:gmklasy}) to this shows a maximum singularity of
\begin{equation}
\left|\Gamma_{{2},{l}}<\partial_v\tilde{\psi},\kappa\cdot
{h^{\ast}_{1,j_1-l-2}}>
\right|\sim
\left(\frac{1}{\gamma}\right)^{1+J_{2,l}}
\left(\frac{1}{\gamma}\right)^{3+J_{1,j_1-l-2}}=
\left(\frac{1}{\gamma}\right)^{5j_1-1}
\label{eq:ex3}
\end{equation}
which is consistent with (\ref{eq:pjasy}).
\end{enumerate}
This completes the proof.
\end{enumerate}

{\bf $\Box$}\end{quote}

%-----------------------------------------------------------
\subsection{Implications of the main result}\label{subsec:implications}
%-----------------------------------------------------------

Theorem \ref{thm:main} allows us to evaluate the
$\gamma\rightarrow0^+$ limit of the theory. Rewriting the amplitude equation
(\ref{eq:modeeqn}) using polar variables $A(t)=\rho(t)\,\exp(-i\theta(t))$, and
the scaling
\begin{equation}
\rho(t)=\gamma^{5/2}\,r(\gamma t)\label{eq:generic}
\end{equation}
from Section \ref{subsec:cubic}, yields
\begin{eqnarray}
\frac{dr}{d\tau}&=& r +
\sum^\infty_{j=1}\left[\gamma^{5j-1}{\rm Re}( p_{j})\right]
r^{2j+1} \label{eq:modeeqa}\\
\frac{d\theta}{dt}&=& \omega - \gamma\sum^\infty_{j=1}
\left[\gamma^{5j-1}{\rm Im}(p_j)\right] r^{2j},
\label{eq:modeeqb}
\end{eqnarray}
where $\tau=\gamma t$. The result (\ref{eq:pjasy}) on the divergence of the
coefficients $p_j$ ensures that the terms in (\ref{eq:modeeqa}) -
(\ref{eq:modeeqb}) are now finite as $\gamma\rightarrow0^+$.
Introducing the asymptotic coefficients
\begin{eqnarray}
c_j(0)&=& \lim_{{\gamma\rightarrow0^+}} \left[ \gamma^{5j-1}
p_j\right]\hspace{0.5
in}( j\geq 1)
\end{eqnarray}
the amplitude equation for a weakly unstable wave becomes
\begin{equation}
\frac{dr}{d\tau}=r +
\sum^\infty_{j=1}\left[{\rm Re}( c_{j}(0))+{\cal O}(\gamma)\right]
r^{2j+1}.\label{eq:aeqn0}
\end{equation}
Unless ${\rm Re}( c_{j}(0))=0$ in each term, the scaling in (\ref{eq:generic})
is required to obtain a finite theory. The circumstances that could force
$c_1(0)=0$ were discussed in Section \ref{subsec:cubic}, and will be analyzed
more completely in Section \ref{sec:special}.

The electric field is obtained from Poisson's equation $\;(imk)\, E_{mk}(t)=
\sum_s \int^\infty_{-\infty} dv {f^u_{mk}}^{(s)}$ using the form of $f^u_{mk}$
in (\ref{eq:fufc})
\begin{equation}
E_{mk}(t)=
\left\{
\begin{array}{cc}
{i}A(1+\sigma\Gamma_{{1}}(\sigma))/{k}&\hspace{0.3in}m=1\\
&\\
{i}A^m\;\Gamma_{{m}}(\sigma)/{mk}&\hspace{0.3in}m>1.\end{array}\right.
\end{equation}
The small growth rate behavior of $\Gamma_m(\sigma)$ can be inferred from
(\ref{eq:gmklasy}) and the expansion
$\Gamma_m(\sigma)=\sum_j\Gamma_{m,j}\sigma^j$. Using
$\sigma^j=\gamma^{5j}r^{2j}$, we find an expansion of $\Gamma_m$ with
a non-singular limit
\begin{equation}
\gamma^{2m-2+5\delta_{m,1}}\Gamma_m=\sum^\infty_{j=0} \gamma^{J_{m,j}+1}
\Gamma_{m,j} r^{2,j},
\end{equation}
and this motivates our definition of $\Gamma^c_m$
\begin{equation}
\Gamma^c_m=\lim_{\gamma\rightarrow0^+}\gamma^{2m-2+5\delta_{m,1}}\Gamma_m=
\sum^\infty_{j=0} b_{m,j}(0) r^{2,j}\label{eq:gmkcrit}
\end{equation}
where
\begin{equation}
\\
b_{m,j}(0)\equiv\lim_{{\gamma\rightarrow0^+}} \left[ \gamma^{J_{m,j}+1}
\Gamma_{m,j}\right].
\end{equation}
With this notation, the asymptotic form of the Fourier expansion
$E(x,t)=\sum_mE_{mk}(t)\exp(imkx)$
of the field can be evaluated
\begin{eqnarray}
E(x,t)&=&\frac{i\gamma^{5/2}}{k}\left[\rule{0in}{0.35in}r(\tau)\,
\left(1+\Gamma^c_1r(\tau)^4+{\cal{O}}(\gamma)\right)
e^{i(kx-\theta(t))}\right.
\label{efield}\\
&&\hspace{0.7in}+\left.\gamma
\sum_{m=2}^{\infty}\;
\frac{\left(\Gamma^c_m\,r(\tau)^m+{\cal{O}}(\gamma)\right)}
{m}e^{im(kx-\theta(t))}\right]+cc.\nonumber
\end{eqnarray}
The asymptotic form may be rewritten more simply as
\begin{equation}
-\frac{iE(x,t)}{\gamma^{5/2}}=\left[
\frac{r(\tau)}{k}\,
\left(1+\Gamma^c_1r(\tau)^4\right)
e^{i(kx-\theta(t))}+ \mbox{\rm cc}\right]+{\cal{O}}(\gamma);
\end{equation}
this describes an electric field comprised of a single Fourier component
whose coefficient scales like $\gamma^{5/2}$ and evolves nonlinearly according
to the asymptotic amplitude equation (\ref{eq:aeqn0}).
The $\gamma^{5/2}$ scaling yields a dramatically smaller field than would
be expected from the trapping scaling $\gamma^{2}$ characteristic of
a model with infinitely massive ions.

An analogous discussion of the asymptotic distribution function $f(x,v,t)$
is substantially more complicated because of singularities in $v$ that
emerge as $\gamma\rightarrow0^+$. This issue is briefly treated in Section
\ref{sec:disc}, but we defer a full analysis to a companion
paper.\cite{jdcajnext}

%_____________________________________________________________________

\section{Special cases}\label{sec:special}
%____________________________________________________________________

Section \ref{subsec:special} identified three settings for which $c_1(0)=0$ and
the $\gamma^{-4}$ singularity of $p_1$ is replaced by a $\gamma^{-3}$
divergence:
\begin{enumerate}
\renewcommand{\theenumi}{\alph{enumi}}
\item Infinitely massive ions: $\kappa^{(s)}=0$ for all $s\neq e$,
\item Zero slope for the resonant ions: $\eta_k^{(s)}(v_p(0))=0$ for all
$s\neq e$,
\item An electron-positron plasma: ${\kappa^{(p)}}^2=1$ for
positrons ($s=p$).
\end{enumerate}
{}From the partial fraction expansion in Lemma \ref{lem:index},
we found the general
condition required to reduce the singularity of a standard integral,
\begin{equation}
\int_{-\infty}^{\infty}\,dv\,{\cal G}(v)=
\int_{-\infty}^{\infty}\,dv\,D_m(\beta,v)^\ast\,D_n(\alpha,v)\,
\sum_s\;(\kappa^{(s)})^{m'}\;\frac{\partial^q\;\eta_l^{(s)}}{\partial v^q},
\hspace{0.5in}(mn>0)
\end{equation}
from $\gamma^{-m-n-q+1}$ to $\gamma^{-m-n-q+2}$; namely,
\begin{equation}
{\sum_s}^\prime\kappa^{(s)}[(-1)^{m'}+(\kappa^{(s)})^{m'-1}]\;
\eta^{(s)}_{k}(v_p(0))=0
\label{eq:special2}
\end{equation}
where the primed sum omits the electrons. The specific instance of
(\ref{eq:special2})
that yields $c_1(0)=0$ corresponds to $m'=3$.

The validity of this general condition varies between the three settings listed
above. For both (a) and (b), (\ref{eq:special2}) holds quite generally
for all integers $m'\geq0$; however for  the third setting (c), the condition
reduces to
\begin{equation}
[(-1)^{m'}+1]\;\eta^{(p)}_{k}(v_p(0))=0,
\label{eq:special3}
\end{equation}
and this will be generally satisfied only for $m'$ odd.

%-----------------------------------------------------------
\subsection{Cases (a) and (b)}
%-----------------------------------------------------------

It is relatively simple to adapt Theorem \ref{thm:main} to accomodate the first
two settings where (\ref{eq:special2}) holds uniformly. The following result
restates Theorem IV.1 of (I) under more general assumptions (note that
the definition of the index in this paper differs from (I)).\cite{jdc95}

\begin{theorem}\label{thm:special} Assume that {\rm (\ref{eq:special2})} holds
for all $m'\geq0$; then
for $j\geq1$, the coefficients in the
expansion of the amplitude equation \mbox{\rm (\ref{eq:modeeqn})} satisfy
\begin{equation}
\lim_{{\gamma\rightarrow0^+}}\;
\gamma^{4j-1}\;\left|p_j\right|<\infty.\label{eq:scpjasy}
\end{equation}
Let $J_{m,j}\equiv(2m+5j-3)+4\delta_{m,0}+5\delta_{m,1}$, then for $j\geq0$,
$m\geq0$ and $m'\geq0$, the indices of $I^{(s)}_{m,j}$ and
$h^{(s)}_{m,j}$ obey
\begin{eqnarray}
{\mbox{\rm Ind }} \left[\sum_s(\kappa^{(s)})^{m'}I^{(s)}_{m,j}\right]&\leq&
J_{m,j}-j-\delta_{m,1}.\label{eq:sckeyind}\\
{\mbox{\rm Ind }} \left[\sum_s(\kappa^{(s)})^{m'}h^{(s)}_{m,j}\right]&\leq&
J_{m,j}+1-j-\delta_{m,1},\label{eq:schind}
\end{eqnarray}
and the integrals in
{\rm (\ref{eq:gmkj})} and {\rm (\ref{eq:hmjint})} satisfy
\begin{eqnarray}
\lim_{{\gamma\rightarrow0^+}}\; \gamma^{J_{m,j}-j}\; \left|\Gamma_{m,j}
\right|&<&\infty\label{eq:scgmklasy}\\
\lim_{{\gamma\rightarrow0^+}}\;\gamma^{J_{m,j}+2-j-\delta_{m,1}}\;
\left|<\partial_v\tilde{\psi},\kappa\cdot h_{m,j}>\right|
&<&\infty.\label{eq:scintest}
\end{eqnarray}
\end{theorem}
The proof is simpler since (\ref{eq:schind}) - (\ref{eq:scintest}) follow
directly from (\ref{eq:sckeyind}) and
and there is no need for special estimates such as (\ref{eq:m=1}) in the $m=1$
case.

\noindent { {\bf Proof.}}
\begin{quote}
\begin{enumerate}
\item The induction proof follows that for Theorem \ref{thm:main}, and we only
highlight the new features brought in by (\ref{eq:special2}).
The recursion relations for $I_{m,j}$ are finite sums; each term of which
has the form $q(\gamma)\,{\cal G}(v)$ where ${\mbox{\rm Ind }} \left[{\cal
G}(v)\right]\geq1$ and the coefficient $q(\gamma)$ may be singular
as ${\gamma\rightarrow0^+}$. According to Lemma \ref{lem:index},
with (\ref{eq:special2}) in force,
if $J={\mbox{\rm Ind }} \left[q(\gamma){\cal G}(v)\right]$, then
$\gamma^{J-1}q(\gamma)\int\,dv\,{\cal G}(v)$ remains finite as
${\gamma\rightarrow0^+}$.

\item We have already shown that (\ref{eq:special2}) implies (\ref{eq:scpjasy})
for $j=1$
so the theorem has been proved for the first level of Table I. As before, we
assume that  (\ref{eq:scpjasy}) - (\ref{eq:scintest}) are valid down to some
arbitrary level of Table I,
and consider what the recursion
relations imply for the coefficients evaluated at the next level. Again, we use
$(m_1,j_1)$ to denote subscripts of coefficients such as $I_{m_1,j_1}$,
$h_{m_1,j_1}$ and $p_{j_1}$ that are to be evaluated from lower order
quantities assumed to satisfy
(\ref{eq:scpjasy}) - (\ref{eq:scintest}).  Following the organization of the
previous proof, we first check that (\ref{eq:sckeyind}) for $I^{(s)}_{m_1,j_1}$
implies (\ref{eq:schind}) - (\ref{eq:scintest}) for $h^{(s)}_{m_1,j_1}$,
$\Gamma_{m_1,j_1}$, and $<\partial_v\tilde{\psi},\kappa\cdot h_{m_1,j_1}>$,
respectively. Then (\ref{eq:sckeyind}) is verified from the recursion relation
for $I^{(s)}_{m_1,j_1}$, and finally (\ref{eq:scpjasy}) is shown to follow from
(\ref{eq:schind}) - (\ref{eq:scintest}).

\item {\bf (Part 1)}: Assume that (\ref{eq:sckeyind}) holds for
$I^{(s)}_{m_1,j_1}$:
\begin{equation}
{\mbox{\rm Ind }} \left[\sum_s(\kappa^{(s)})^{m'}I^{(s)}_{m_1,j_1}\right]\leq
J_{m_1,j_1}-j_1-\delta_{m_1,1},\label{eq:sckeyind2}
\end{equation}
then the  properties (\ref{eq:schind}) - (\ref{eq:scintest})
are easily obtained as follows.
First, the integrand of $\Gamma_{m_1,j_1}$ in (\ref{eq:gmkj})
has index
\begin{equation}
{\mbox{\rm Ind }}
\left[\sum_s\,{I^{(s)}_{m_1,j_1}}/(v-z_{m,j})\right]
\leq J_{m_1,j_1}-j_1-\delta_{m_1,1}+1\label{eq:intind}
\end{equation}
from (\ref{eq:sckeyind}) so the expression
\begin{equation}
\gamma^{J_{m_1,j_1}-j_1-\delta_{m_1,1}}
\int^\infty_{-\infty}\,dv\,\frac{\sum_s
I^{(s)}_{m,j}(v)}{v-z_{m,j}},\label{eq:finite}
\end{equation}
remains finite as ${\gamma\rightarrow0^+}$; this proves (\ref{eq:scgmklasy})
since the $\delta_{m_1,1}$ in the exponent compensates for the added
singularity of
$\Lambda_{k}(z_{1,j_1})$ in (\ref{eq:gmkj}).  Next, (\ref{eq:schind}) follows
by applying (\ref{eq:sckeyind2}) and  (\ref{eq:scgmklasy}) to the identity
(\ref{eq:hmj}).
 Finally, the integrand in (\ref{eq:scintest}) has index
\begin{equation}
{\mbox{\rm Ind }}
\left[\sum_s\,\partial_v{\tilde{\psi}^{(s)}}\kappa^{(s)} h^{(s)}_{m,j}\right]
\leq J_{m_1,j_1}-j_1-\delta_{m_1,1}+3\label{eq:intind2}
\end{equation}
from (\ref{eq:schind}) so the bound in (\ref{eq:scintest}) follows from
(\ref{eq:special2}) and Lemma \ref{lem:index}.

\item {\bf (Part 2)}: {\em Verification of {\rm(\ref{eq:sckeyind2})} - the
index of $I_{m_1,j_1}$:}\\
The consequences of (\ref{eq:special2}) for the index calculation are
incorporated in (\ref{eq:scpjasy}) - (\ref{eq:scgmklasy}).
The index of each term on the right hand side of the
recursion relations for $I^{(s)}_{m_1,j_1}(v)$ can be evaluated by applying
these conclusions regarding the lower order coefficients; this is done exactly
as in the proof of (\ref{eq:keyind2}) in Theorem \ref{thm:main}.
The bound in (\ref{eq:sckeyind2}) follows as before.

\item {\bf (Part 3)}: {\em Verification of {\rm(\ref{eq:scpjasy})} for
$p_{j_1}$}\\
It is straightforward to check that application of (\ref{eq:schind}) -
(\ref{eq:scintest}) to
the right hand side of the recursion relations (\ref{eq:pj}) - (\ref{eq:bp})
yields the estimate in
(\ref{eq:scpjasy}) just as in the proof of Theorem \ref{thm:main}.

\end{enumerate}

{\bf $\Box$}\end{quote}

In light of the discussion in Section \ref{subsec:implications}, the
theorem tells us that the scaling in (\ref{eq:generic}) is replaced by
\begin{equation}
\rho(t)=\gamma^{2}\,r(\gamma t)\label{eq:trapping}
\end{equation}
for the special cases (a) and (b). This modification in turn implies an
electric field characterized by the ``trapping scaling'' $E_k\sim\gamma^{2}$;
the discussion in (I) provides detailed expressions for
the asymptotic form of $E(x,t)$ and $f^{(e)}(x,v,t)$ in the limit
$\gamma\rightarrow0^+$ when the ions are fixed.\cite{jdc95}

%-----------------------------------------------------------
\subsection{An electron-positron plasma}\label{subsec:ep}
%-----------------------------------------------------------

The result $p_j\sim\gamma^{-4j+1}$ for the two special cases just considered
is not correct for the electron-positron system with
$\kappa^{(p)}=-\kappa^{(e)}=1$. Although $p_1\sim\gamma^{-3}$ does hold, we
find at fifth order $p_2\sim\gamma^{-8}$  instead of $p_2\sim\gamma^{-7}$.
Thus the trapping scaling of the amplitude in (\ref{eq:trapping}) is not
sufficient to obtain a finite expansion.
We  summarize the evaluation of $p_2$
from (\ref{eq:pj}) - (\ref{eq:bp}), but leave the comprehensive
analysis of this case as an open problem.

The exact recursion relation for $p_2$ is
\begin{eqnarray}
\lefteqn{-ikp_2=-<\partial_v\tilde{\psi},\kappa\cdot  (h_{0,1}-h_{2,1})>
-\frac{1}{2}<\partial_v\tilde{\psi}, \kappa\cdot
{\psi}^\ast>\Gamma_{{2},{1}}}
\label{eq:p2sc}\\
&&\hspace{1.5 in}
-<\partial_v\tilde{\psi},\kappa\cdot
h_{0,0}>\Gamma_{{1},{0}}
+<\partial_v\tilde{\psi},\kappa\cdot  h_{2,0}>\Gamma^\ast_{{1},{0}}\nonumber\\
&&\hspace{0.5 in}-
\left\{\frac{1}{2}<\partial_v\tilde{\psi}, \kappa\cdot  {h^{\ast}_{1,0}}>
\Gamma_{{2},{0}}+\frac{\Gamma_{{3},{0}}}{3}
<\partial_v\tilde{\psi}, \kappa\cdot  {h^{\ast}_{2,0}}>
-\frac{\Gamma^\ast_{{2},{0}}}{2}
<\partial_v\tilde{\psi},\kappa\cdot  h_{3,0}>\right\}, \nonumber
\end{eqnarray}
and, in the bracketed terms, $\Gamma_{2,0}$ and $\Gamma_{3,0}$ are non-singular
so Theorem \ref{thm:main} ensures the divergence of these terms cannot exceed
$\gamma^{-4}$; hence they are subdominant:
\begin{eqnarray}
\lefteqn{-ikp_2=-<\partial_v\tilde{\psi},\kappa\cdot  (h_{0,1}-h_{2,1})>
-\frac{1}{2}<\partial_v\tilde{\psi}, \kappa\cdot
{\psi}^\ast>\Gamma_{{2},{1}}}
\label{eq:p2sc2}\\
&&\hspace{1.5 in}
-<\partial_v\tilde{\psi},\kappa\cdot
h_{0,0}>\Gamma_{{1},{0}}
+<\partial_v\tilde{\psi},\kappa\cdot  h_{2,0}>\Gamma^\ast_{{1},{0}}
+{\cal O}(\gamma^{-4}).\nonumber
\end{eqnarray}
Here consider the two terms involving $\Gamma_{{1},{0}}$. In the general case,
Theorem \ref{thm:main} gives $\Gamma_{{1},{0}}\sim\gamma^{-5}$, but
this drops to $\gamma^{-4}$ when we take
(\ref{eq:special2}) into account for $m'$
odd. The
integral $<\partial_v\tilde{\psi},\kappa\cdot  h_{2,0}>$ is non-singular
so $<\partial_v\tilde{\psi},\kappa\cdot
h_{2,0}>\Gamma^\ast_{{1},{0}}\sim\gamma^{-4}$ and we neglect this term also.
Next $<\partial_v\tilde{\psi},\kappa\cdot
h_{0,0}>$ is written out to find
\begin{equation}
<\partial_v\tilde{\psi},\kappa\cdot
h_{0,0}>=\frac{i}{2k\gamma}<\partial_v\tilde{\psi},\kappa^2\cdot
\partial_v(\psi^\ast-\psi)>;
\end{equation}
this integrand has index equal to 3, but contains an odd power of $\kappa$ so
(\ref{eq:special2}) holds and the
divergence cannot exceed $\gamma^{-2}$. Overall,
$<\partial_v\tilde{\psi},\kappa\cdot
h_{0,0}>\Gamma_{{1},{0}}$ has  a maximum singularity of  $\gamma^{-7}$.
A careful
examination of the first term in (\ref{eq:p2sc2}) yields the same estimate
$<\partial_v\tilde{\psi},\kappa\cdot  (h_{0,1}-h_{2,1})>\sim\gamma^{-7}$.

The dominant singularity
in $p_2$ arises from the remaining term
\begin{equation}
p_2=-\frac{i\Gamma_{{2},{1}}}{2k}
<\partial_v\tilde{\psi}, \kappa\cdot{\psi}^\ast>+{\cal O}(\gamma^{-7}),
\label{eq:p2dominant}
\end{equation}
and from (\ref{eq:close}) and (\ref{eq:limit2}), we obtain
\begin{equation}
<\partial_v\tilde{\psi}, \kappa\cdot{\psi}^\ast>=
-\left(\frac{k}{\gamma}\right)^2
\left[\frac{i\,\pi\,\eta_k^{(p)}(v_p(0))}
{\Lambda_k'(v_p+i0)}+{\cal O}(\gamma)\right].
\end{equation}
The evaluation of $\Gamma_{{2},{1}}$ starts from (\ref{eq:gmkj})
\begin{eqnarray}
\Gamma_{{2},{1}}&=&\frac{-i/2k}{\Lambda_{2k}(z_{{2},{1}})}\sum_s
\int^\infty_{-\infty}\,dv\,\frac{I^{(s)}_{{2},{1}}(v)}{v-z_{{2},{1}}},
\label{eq:gmkj21}
\end{eqnarray}
and the recursion relation (\ref{eq:I2l}) for $I_{{2},{1}}$
\begin{eqnarray}
I_{2,1}(v)=-2p_{1}\,h_{2,0}
+\frac{i}{k}\,\frac{\partial}{\partial v}\kappa\cdot\left\{
h_{1,0}+\psi \Gamma_{{1},{0}} -h_{3,0}
+\frac{1}{3} {\psi}^\ast \Gamma_{{3},{0}}+ \frac{1}{2}
h_{0,0} \Gamma_{{2},{0}} \right\}.\label{eq:I21}
\end{eqnarray}
Note that $\Lambda_{2k}(z_{{2},{1}})=3/4 +{\cal O}(\gamma)$ from
(\ref{eq:specfcnid}) so any singularities arise from the integral
\begin{eqnarray}
\int^\infty_{-\infty}\,dv\,\frac{\sum_s I^{(s)}_{{2},{1}}(v)}{v-z_{{2},{1}}}
&=&-2p_{1}\int^\infty_{-\infty}\,dv\,
\frac{\sum_s h^{(s)}_{2,0}(v)}{v-z_{{2},{1}}}\label{eq:I21int}\\
&&+\frac{i}{k}\,\int^\infty_{-\infty}
\frac{\,dv\,}
{(v-z_{{2},{1}})^2}
\sum_s\kappa^{(s)}
\left[h^{(s)}_{1,0}+\psi^{(s)} \Gamma_{{1},{0}} -h^{(s)}_{3,0}
+\frac{1}{3} {{\psi}^{(s)}}^\ast \Gamma_{{3},{0}}
+\frac{1}{2} h^{(s)}_{0,0} \Gamma_{{2},{0}}\right].\nonumber
\end{eqnarray}
The strongest singularity here occurs in the integral over $h^{(s)}_{1,0}$,
and all other
terms in (\ref{eq:I21int}) are subdominant; thus we can focus attention on
the expression
\begin{equation}
\int^\infty_{-\infty}\,dv\,\frac{\sum_s I^{(s)}_{{2},{1}}(v)}{v-z_{{2},{1}}}
=\frac{i}{k}\,\int^\infty_{-\infty}\,dv\,
\frac{\sum_s\kappa^{(s)}h^{(s)}_{1,0}}
{(v-z_{{2},{1}})^2}
+{\cal O}(\gamma^{-4}).
\end{equation}
When $h^{(s)}_{1,0}$ is evaluated using (\ref{eq:hmj}) we obtain
\begin{equation}
\int^\infty_{-\infty}\,dv\,\frac{\sum_s I^{(s)}_{{2},{1}}(v)}{v-z_{{2},{1}}}
=\frac{1}{k^2}\int^\infty_{-\infty}\,dv\,
\frac{\sum_s\kappa^{(s)}I^{(s)}_{1,0}}
{(v-z_{{2},{1}})^2(v-z_{{1},{0}})}
+{\cal O}(\gamma^{-4}).
\end{equation}
Now the recursion relation (\ref{eq:I1l}) for $I^{(s)}_{1,0}$ is applied, and
examination
of the various terms shows that the dominant singularity arises from
the occurrence of $\psi^\ast$ in $h_{0,0}$ so that (\ref{eq:gmkj21}) can be
reduced to
\begin{eqnarray}
\Gamma_{{2},{1}}&=&\left(\frac{i\left[1+{\cal O}(\gamma)\right]}
{3\gamma k^5}\right)
\int^\infty_{-\infty}\,dv\,
\frac{\sum_s(\kappa^{(s)})^3\partial^2_v{\psi^{(s)}(v)}^\ast}
{(v-z_{{2},{1}})^2(v-z_{{1},{0}})}
+{\cal O}(\gamma^{-5}).
\label{eq:gmkj21int}
\end{eqnarray}
Finally a partial fraction expansion of the remaining integral yields a
$\gamma^{-5}$ singularity
\begin{equation}
\int^\infty_{-\infty}\,dv\,
\frac{\sum_s(\kappa^{(s)})^3\partial^2_v{\psi^{(s)}(v)}^\ast}
{(v-z_{{2},{1}})^2(v-z_{{1},{0}})}=\frac{1}{\gamma^5}
\left[-\left(\frac{\pi }{8}\right)k^5\eta_k^{(p)}(v_p(0))+{\cal
O}(\gamma)\right],
\end{equation}
and leads to the asymptotic form of $p_2$ from (\ref{eq:p2dominant})
\begin{equation}
p_2=\frac{1}{\gamma^8}
\left[\left(\frac{i\pi^2k}{48}\right)
\frac{\eta_k^{(p)}(v_p)^2}{\Lambda_k'(v_p+i0)}+{\cal O}(\gamma)\right]
+{\cal O}(\gamma^{-7}).
\label{eq:p2final}
\end{equation}

Since ${\rm Im}\Lambda_k'(v_p+i0)$ is
typically non-zero, this shows that ${\rm Re}(p_2)\sim \gamma^{-8}$
and implies that  the trapping scaling (\ref{eq:trapping}) leaves a residual
singularity
at fifth order. With $\rho(t)=\gamma^\beta\,r(\gamma t)$ the rescaled equation
becomes
\begin{equation}
\frac{dr}{d\tau}=r+\gamma^{2\beta-1}{\rm Re}(p_1)\,r^3+\gamma^{4\beta-1}{\rm
Re}(p_2)
\,r^5+\cdots
\end{equation}
so $\beta\geq9/4$ is required to overcome the $\gamma^{-8}$ divergence in
$p_2$.

\section{Discussion}\label{sec:disc}

 Our analysis of an unstable electrostatic wave expresses the Vlasov
distribution function $f(x,v,t)$ in terms of the wave amplitude, i.e.
\begin{equation}
f(x,v,t)=\left[A(t)e^{ikx}\psi(v) +  \;\;cc\right]+\sum_{m=-\infty}^\infty
e^{imkx}\,H_{mk}(v,A,A^\ast)\label{eq:dist}
\end{equation}
where $A(t)$ evolves according to an amplitude equation,
\begin{equation}
\dot{A}(t)=A\,p(|A^2|).\label{eq:aeqndisc}
\end{equation}
In the small growth rate limit, the eigenvalue of the
 unstable mode approaches the imaginary axis and, at $\gamma=0$,
merges with a neutrally stable continuous spectrum on the
axis.\cite{jdc95,cra1} This ``interaction'' between the unstable
mode and the neutrally stable continuum leads to singular features in
the asymptotic form of the amplitude equation (\ref{eq:aeqndisc}).\cite{cra4}

These singularities are physically significant since they reflect the
very strong nonlinear interaction between the wave and the resonant particles
at the linear phase velocity. Remarkably, the quantitative form of the
singularities can be consistently interpreted as fixing overall scalings
of the nonlinear solution with the linear growth rate in the asymptotic
limit $\gamma\rightarrow0^+$. This interpretation is achieved through
a singular
transformation $|A(t)|=\gamma^\beta\,r(\gamma t)$ of the amplitude in
which the exponent $\beta$ is fixed by the criterion that the rescaled
dynamics,
\begin{equation}
\frac{dr}{d\tau}=r(\tau)\frac{{\rm Re}(p(\gamma^{2\beta}r(\tau)^2))}
{\gamma},\label{eq:reqndisc}
\end{equation}
is free of singularities as $\gamma\rightarrow0^+$.
Our conclusions are summarized in Table II; with the notable exception
of the electron-positron system, we can rigorously establish an exact value
of $\beta=5/2$ in the typical instability and $\beta=2$ for certain exceptional
models. In the electron-positron case, the we have fixed only
a range of possible $\beta$ values from the explicit calculation of $p(|A|^2)$
through fourth order in $|A|$; the complexity of the recursion
relations in this case prevents the iterative determination of the
dominant singularities at each order.

The shift from $\beta=2$ to $\beta=5/2$
as the true asymptotic scaling due to finite mass ions at the phase velocity
of the wave is our most significant conclusion regarding the physical
character of the nonlinear wave. Since previous studies of the
asymptotic scaling of single wave instabilities have focussed primarily on
plasma waves and assumed fixed ions, this shift was not anticipated and
was discovered purely as a consequence of the singularities in the
amplitude equation (\ref{eq:aeqndisc}).  Numerical simulations of unstable
ion-acoustic waves or unstable plasma waves in systems with finite mass ratios
would be of considerable interest in this connection. In particular, for
plasma waves one expects a crossover in the asymptotic scaling from the
trapping scaling ($\beta=2$) to $\beta=5/2$ at sufficiently small growth
rates. Predictions of this crossover have been made elsewhere based on the
competition between the $\gamma^{-4}$ and $\gamma^{-3}$ singularities in the
cubic coefficient.\cite{jdcaj}

In this paper we have not analyzed the asymptotic form of $f(x,v,t)$,
but it is clear from the expansions for $H(x,v,A,A^\ast)$ that this is an
interesting question. For example, consider the leading coefficient in $h_0(v)$
from (\ref{eq:h00}),
\begin{equation}
h_{0,0}(v)= -\frac{1}{k^2}\frac{\partial}{\partial v}
\left[\frac{\kappa^2\cdot\eta_k}{(v-z_0)(v-z^\ast_0)} \right];
\label{eq:h00f}
\end{equation}
there are poles in the complex-velocity plane at $z_0$ and $z^\ast_0$
that approach the real velocity axis at $v_p$ in the limit
$\gamma\rightarrow0^+$. Thus the coefficient becomes a singular function
at the phase velocity when $\gamma=0$;
on the other hand, for $v\neq v_p$, the
coefficient is well-behaved at small growth rates. In the
distribution function (\ref{eq:dist}) this particular coefficient is multiplied
by $|A|^2$ so that after rescaling the amplitude we have a term of
the form $r^2\gamma^{2\beta}h_{0,0}(v)$.
Now the emerging singularity in velocity
space is weighted by a factor of $\gamma^{2\beta}$,
and one would like to understand
whether the resulting product is finite or singular at $\gamma=0$.  A detailed
analysis of this problem will be presented in a forthcoming paper, but the
primary conclusion can be simply summarized.\cite{jdcajnext} Inside a
neighborhood of width ${\cal O}(\gamma)$, and centered on $v=v_p$, the
singularities in the coefficients of $H$ are balanced by the factors of
$\gamma$
introduced by the rescaling of $|A|$ and one can obtain finite, non-zero limits
for the weighted coefficients. Outside this small neighborhood of the
phase velocity, the factors of $\gamma$ force the weighted coefficients to
vanish at $\gamma=0$.

The connection between coefficient singularities and nonlinear scalings
that is studied in this paper is relevant in other settings. As mentioned
above, a key feature of our
instabilities is the emergence of the critical eigenvalues from a neutrally
stable continuous spectrum. It has been appreciated recently that this
characteristic
is shared by a diverse set of additional examples which include the onset of
linear instability in fluid shear flows\cite{case2}-\cite{mas}, instabilities
of
solitary waves\cite{pegowein1}-\cite{pegwein3},
synchronization of large systems of coupled
oscillators,\cite{sm}
and bifurcations in
``mean field'' descriptions of the dynamics of bubble clouds in
fluids\cite{russo}. In the phase-dynamical models of coupled oscillators,
the amplitude equations of the weakly
unstable modes have been analyzed in the $\gamma\rightarrow0^+$
limit.\cite{jdc0,jdc3} In these studies, the coefficients can be
singular or non-singular depending on how the oscillators are coupled, but when
singularities occur in the limit of small growth they can be absorbed by
singular rescalings of the amplitude in a procedure analogous to the method
used in this paper. The specific values of the scaling exponents are different,
but yield predictions for the emergence of the nonlinear synchronized state.

It seems likely that the analysis of singular amplitude equations  provides
a general means to determine  scaling exponents for weakly unstable systems.
The
exponents in turn provide a quantitative measure of the strong nonlinear
interaction between the neutrally stable modes and the weakly unstable modes.

\acknowledgments
This work supported by NSF grant PHY-9423583.

\widetext
\begin{table}\label{table:flow}
\caption{Order of calculation of $h_{m,j}(v)$ and $p_j$ from $\psi(v)$. The
flow of calculation of the $h_{m,j}(v)$ is indicated
by moving downward. From $\psi(v)$, $h_{0,0}$ and $h_{2,0}$ can be calculated
 and then $p_1$ determined; $h_{1,0}$ and $h_{3,0}$ are calculated next from
$\{p_1, h_{0,0}, h_{2,0}\}$ and then $h_{0,1}$ and $h_{2,1}$ can be evaluated.
This then determines $p_2$, and so forth. For $j_1\geq2$, $p_{j_1}$ requires
prior
calculation of $h_{m,j}$ for $0\leq m\leq {j_1}+1$ and $0\leq j\leq {j_1}-m+1-2
(\delta_{m,0}+\delta_{m,1}).$}
\begin{tabular}{l|cccccccc}
   &$ m=0$&$m=1$&$m=2$&$m=3$&$m=4$&$m=5$&$m=6$&$\cdots$\\ \hline
\\
$p_0$&  & $\psi(v)$   &   & & & & & \\
\hline
\\
$p_1$&$h_{0,0}$&  -         &$h_{2,0}$& & & & &\\
\hline
     &            &$h_{1,0}$&            &$h_{3,0}$& & & &\\
$p_2$&$h_{0,1}$&            &$h_{2,1}$&            & & & & \\
\hline

     &            &$h_{1,1}$&            &            &$h_{4,0}$& & & \\
$p_3$&$h_{0,2}$&            &$h_{2,2}$&$h_{3,1}$&            & & &\\
\hline
     &            &$h_{1,2}$&            &         &      &$h_{5,0}$& &\\
$p_4$&$h_{0,3}$&            &$h_{2,3}$& $h_{3,2}$&$h_{4,1}$ && &\\
\hline
     &            &$h_{1,3}$&            &         &      &&$h_{6,0}$ &\\
$p_5$&$h_{0,4}$&            &$h_{2,4}$& $h_{3,3}$&$h_{4,2}$ &$h_{5,1}$& &\\
\hline
     &            &&            &         &      && &\\
$\vdots$&&            && & && &\\
\end{tabular}
\label{table1}
\end{table}

\begin{table}
\caption{Summary of the exponent $\beta$ obtained from the amplitude
equation. The cases (a), (b) and (c) are the exceptional cases discussed in
Section \ref{sec:special}.}
\vspace{0.25in}
\begin{tabular}{lcc}
System&$\beta$&Comment\\ \hline
\\
\hspace{0.25in}$c_1(0)\neq0$&  $\frac{5}{2}$ & generic result for multiple
mobile species\\ \hline
\\
(a) $c_1(0)=0$&  2 &  infinitely massive ions\\ \hline
\\
(b) $c_1(0)=0$&  2 & $\eta_k^{(s)}(v_p)=0$ for each ion species\\ \hline
\\
(c) $c_1(0)=0$ & $\frac{9}{4}\leq\beta\leq\frac{5}{2}$ & an electron-positron
plasma\\ \hline
\\
\end{tabular}
\label{table2}
\end{table}

\end{document}